\documentclass[lettersize,journal]{IEEEtran}
\usepackage{amsmath,amsfonts}
\usepackage{algorithm}
\usepackage{array}
\usepackage[caption=false,font=normalsize,labelfont=sf,textfont=sf]{subfig}
\usepackage{textcomp}
\usepackage{stfloats}
\usepackage{url}
\usepackage{verbatim}
\usepackage{graphicx}

\usepackage{cite}
\usepackage[colorlinks,
linkcolor=blue,   
anchorcolor=blue, 
citecolor=blue,      
]{hyperref}

\usepackage{bbding}
\usepackage{makecell}
\usepackage{color}
\usepackage{booktabs}
\usepackage{makecell}
\usepackage{array,multirow}
\usepackage{tabularx}
\graphicspath{{./Figs/}}
\usepackage{algpseudocode}
\usepackage{makecell}
\begin{document}

\title{FACE: Few-shot Adapter with Cross-view Fusion for Cross-subject EEG Emotion Recognition}

\author{Haiqi Liu~\IEEEmembership{Student Member,~IEEE}, C. L. Philip Chen~\IEEEmembership{Life Fellow,~IEEE}, Tong Zhang~\IEEEmembership{Senior Member,~IEEE}
	
	\thanks{
	This work was supported in part by the National Natural Science Foundation of China under Grant 62536004, in part by the Key-Area Research and Development Program of Guangdong Province under Grant 2023B0303030001, in part by the Science and Technology Program of Guangzhou under Grant 2024A04J6310, in part by the GJYC Program of Guangzhou under Grant 2025B01Q0001, and in part by the Fundamental Research Funds for the Central Universities under Grant 2026ZYGXZR032.
		\textit{(Corresponding author: Tong Zhang)}
	}
	
	\thanks{The authors are with the Guangdong Provincial Key Laboratory of Computational AI Models and Cognitive Intelligence and the School of Computer Science and Engineering, South China University of Technology, Guangzhou 510006, China; the Pazhou Lab, Guangzhou 510335, China; and the Engineering Research Center of the Ministry of Education on Health Intelligent Perception and Paralleled Digital-Human, Guangzhou, China. (e-mail: tony@scut.edu.cn). }
	
	\thanks{The source code is available at https://github.com/cs-jerry/face.}
}

\markboth{IEEE Transactions on Affective Computing}%
{Shell \MakeLowercase{\textit{et al.}}: A Sample Article Using IEEEtran.cls for IEEE Journals}


\maketitle

\begin{abstract}
	Cross-subject EEG emotion recognition is challenged by significant inter-subject variability and intricately entangled intra-subject variability. Existing works have primarily addressed these challenges through domain adaptation or generalization strategies. However, they typically require extensive target subject data or demonstrate limited generalization performance to unseen subjects. Recent few-shot learning paradigms attempt to address these limitations but often encounter catastrophic overfitting during subject-specific adaptation with limited samples. This article introduces the few-shot adapter with a cross-view fusion method called FACE for cross-subject EEG emotion recognition, which leverages dynamic cross-view fusion and effective subject-specific adaptation. Specifically, FACE incorporates a cross-view fusion module that dynamically integrates global brain connectivity with localized patterns via subject-specific fusion weights to provide complementary emotional information. Moreover, the few-shot adapter module is proposed to enable rapid adaptation for unseen subjects while reducing overfitting by enhancing adapter structures with meta-learning. Experimental results on three public EEG emotion recognition benchmarks demonstrate the effectiveness of FACE under few-shot cross-subject evaluation.
\end{abstract}


\section{Introduction}
\IEEEPARstart{U}{nderstanding} human emotions is fundamental and crucial to advancing fields such as human-computer interaction~\cite{lee2024encoding} and mental health~\cite{karl2024role}. Electroencephalography (EEG) has recently emerged as a remarkable tool for capturing subjects’ neural responses to emotional states~\cite{Li2022ACM}. EEG-based emotion recognition remains challenging due to the substantial inter-subject variance in brain activity patterns~\cite{10453943,10609541,Xu_2025_ICCV}. Additionally, intra-subject variance arises from the non-stationary nature of EEG signals, which exhibit variations in frequency and amplitude over time within the same subject. In most cases, these variances entangle with each other, diminishing the ability of models trained on one group to generalize effectively to unseen subjects.

\begin{figure} [!t]
	\centering
	\includegraphics[scale=0.45]{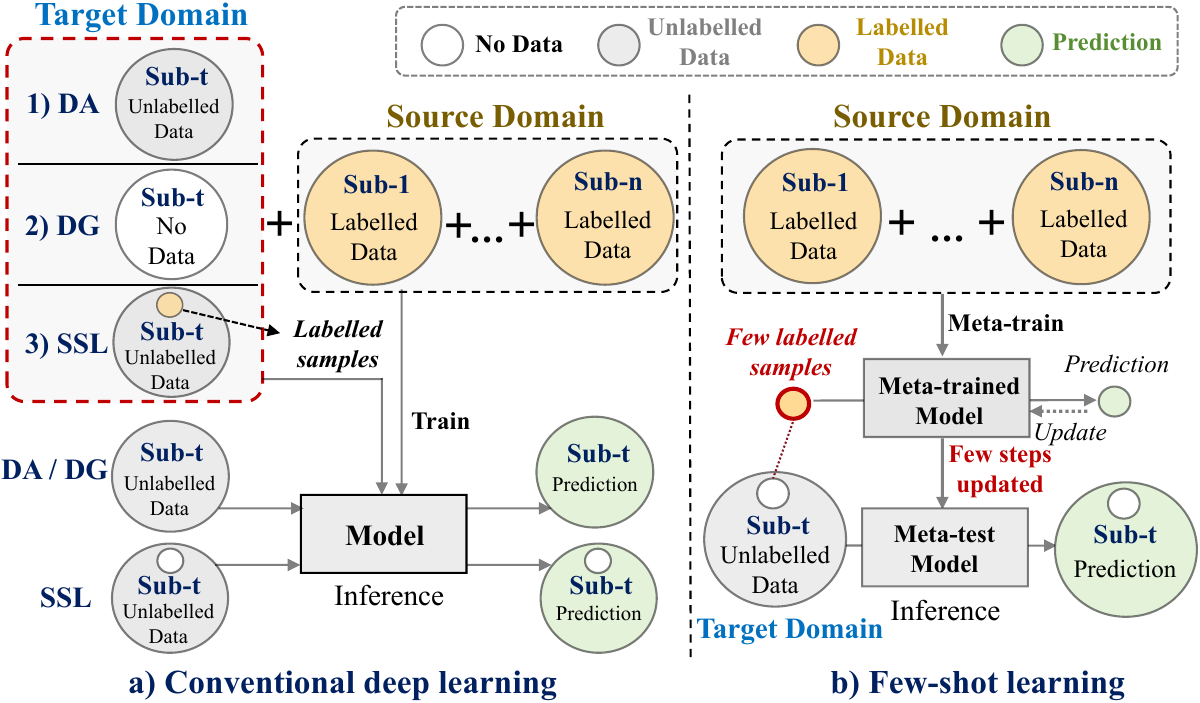}
	\caption{Comparison of training data and processes between Few-Shot Learning (FSL) and traditional deep learning (DL) in cross-subject EEG emotion recognition. Traditional DL primarily relies on additional unlabeled target data for training, whereas FSL learns a subject-agnostic model in the source domain and fine-tunes it using only a few target-domain samples with few steps. DA denotes Domain Adaptation, DG denotes Domain Generalization, and SSL denotes Semi-Supervised Learning.
	}
	\label{fig:task}
\end{figure}

Existing studies are mainly dedicated to addressing the inter-subject variance through domain adaptation (DA) and domain generalization (DG) strategies. DA-based methods learn subject-specific representations by aligning the marginal distributions of source and target subjects~\cite{wilson2020survey}. Although these methods~\cite{9904850,10609541,10839595} have achieved notable results, they require large amounts of unlabeled target subject data during training. This requirement restricts their applicability in scenarios with scarce or non-specific target subject data~\cite{kouw2019review}. In contrast, DG-based methods attempt to learn subject-invariant representations from source data~\cite{robey2021model}. However, they often struggle to generalize to unseen target domains due to their dependence on source data diversity~\cite{khoee2024domain}. Moreover, the absence of target domain data in training makes it challenging for these methods to estimate subtle domain shifts and capture target subject-specific patterns. Recently, few-shot learning~\cite{gharoun2024meta} (FSL) approaches have emerged as a promising alternative to address this issue. These methods~\cite{10822213,Chen_2025, ng2024subject} leverage the FSL method to learn a whole subject-agnostic initialization that can quickly adapt to new subjects with only a few labeled samples (See Fig.~\ref{fig:task}). Nonetheless, the scarcity of labeled EEG data from target subjects often leads to catastrophic overfitting when performing whole model parameter updates with limited training steps. These limitations result in suboptimal performance for novel subjects. Based on these observations, this article encourages a partial parameter update strategy for target subject adaptation, prioritizing rapid adaptation over full model fine-tuning. We introduce the Few-shot Adapter module, which incorporates the adapter structure with the meta-learning paradigm. This design facilitates capturing subject-specific emotional discriminative cues while reducing overfitting under few-shot conditions. 

Furthermore, effectively modeling intra-subject EEG variability continues to be difficult~\cite{10556575}. Earlier studies attempted to mitigate this issue by introducing statistical features such as power spectral density (PSD) and phase-locking value (PLV) to capture complementary information about the EEG signal~\cite{Li2022ACM}. However, these methods suffer from heterogeneous feature spaces, leading to redundant information and noise accumulation~\cite{9760385}. Recent work has shifted toward automatically learning discriminative patterns, leveraging multi-scale learning~\cite{10496191}, contrastive learning~\cite{9765326}, or spatial-temporal attention~\cite{9751142} within graph convolutional networks (GCNs). Although these approaches can reveal complex brain functional connectivity associated with emotion, they typically rely on a single view of information or uniform fusion weights across all unseen subjects, limiting their capacity to fully exploit subject-specific discriminative details. It is reasonable to assume that fusion weights for different views should also be subject-specific because inter-subject variance is inherently coupled with intra-subject variance in EEG signals. To address this gap, this article proposes a cross-view fusion module that dynamically learns a unified representation across views, using subject-specific fusion weights.

To address the above challenges, this article proposes a few-shot adapter with a cross-view fusion  framework named FACE for cross-subject EEG emotion recognition. Departing from existing FSL-based methods in this field, FACE innovatively incorporates complementary spatial view information and encourages subject-specific adaptation from both the feature fusion and adjustment aspects only, thereby achieving superior cross-subject generalization. Specifically, FACE comprises two key components: a cross-view fusion (CVF) module and a few-shot adapter (FSA) module. Recent studies~\cite{pessoa2017network,chen2018domain} have demonstrated that emotional processing depends on both the activity of specific brain regions and the coordination of large-scale brain networks. Building on this insight, the CVF module extends the conventional GCN view by introducing a cross-view architecture. The GCN view extracts global connectivity patterns, while a complementary local view identifies region-specific emotional information. The CVF module projects two view representations into a shared space and performs subject-specific complementary fusion with meta-learning. Moreover, inspired by the adapter architecture in the computer vision field, this article introduces a FSA module to enable rapid calibration for unseen subjects in the EEG emotion recognition task. Due to the significant inter-subject variability in EEG signals, this article incorporates batch normalization layers with each residual adapter layer to statistically regularize inter-subject variability. Moreover, the adapter employs a meta-learning paradigm to achieve more effective calibration. To the best of our knowledge, this is the first work to integrate adapters with meta-learning for few-shot cross-subject EEG emotion recognition. Comprehensive evaluations on three widely used benchmarks demonstrate that FACE outperforms state-of-the-art cross-subject EEG emotion recognition methods.

In summary, this work presents the following contributions:
\begin{enumerate}
	\item This article proposes FACE, a novel method to enhance cross-subject EEG emotion recognition through subject-specific adaptation via joint feature fusion and adjustment. Extensive experiments across three widely used benchmarks demonstrate that FACE achieves  comparable performance to state-of-the-art approaches.
	\item The CVF module effectively combines global brain connectivity and local patterns, utilizing meta-learned cross-view fusion. It dynamically provides a comprehensive subject-specific unified representation. 
	\item The FSA module incorporates an adapter structure and meta-learning paradigm to enable rapid subject-specific emotional feature adjustment for unseen subjects.
\end{enumerate}

The remainder of the article is organized as follows. Section~\ref{related_work} reviews the relevant literature. Section~\ref{method} details the proposed FACE model. Section~\ref{exp_result} presents the experimental results and comparative analysis. Section~\ref{discussion} critically examines the impact of each module and the parameter settings. Finally, Section~\ref{conclusion} summarizes our contributions and discusses future research directions.

\section{Related Work}
\label{related_work}
\subsection{Cross-Subject EEG Emotion Recognition}
Cross-subject EEG emotion recognition has garnered significant attention due to its pivotal role in brain-computer interfaces~\cite{JAFARI2023107450}. The task is particularly challenging due to the inherent non-stationarity of EEG signals and the complexity of human emotional expression. Early approaches focused on handcrafted statistical features combined with simple classifiers, but these yielded limited performance~\cite{houssein2022human}. Recent work has shifted toward deep supervised learning to obtain more expressive EEG representations~\cite{song2020instance,9857970,10496191,10568943}. Song \textit{et al.} introduced an instance-adaptive graph network with an additional branch to model dynamic inter-regional connectivity~\cite{song2020instance}. Sun \textit{et al.} employed a dual-branch framework to capture complementary temporal and spectral cues from different statistical descriptors~\cite{9857970}. Jin \textit{et al.} proposed a pyramidal GCN that leverages multi-scale brain-region structures to improve performance~\cite{10496191}. Gong \textit{et al.} integrated hemispheric asymmetry and spatial–spectral–temporal features to capture extensive emotional cues~\cite{10568943}. Despite these advances, existing methods largely depend on predefined regional priors or handcrafted descriptors, and their fusion mechanisms are typically fixed or based on generic attention schemes. These designs provide limited flexibility in adapting to subject-specific variability, motivating the need for the comprehensive and subject-specific fusion mechanism. More recently, semi-supervised learning paradigms have been applied to EEG emotion recognition to refine individual-specific mappings better. Li \textit{et al.} combined meta-learning and semi-supervised learning to facilitate subject-adaptation representations~\cite{li2023novel}. Guarneros \textit{et al.} proposed semi-supervised multi-source joint distribution adaptation to align distributions across different subjects~\cite{10214058}. However, these methods generally require substantial amounts of unlabeled target-domain data and may suffer from instability arising from multi-objective optimization. To address these issues, this article explores the few-shot cross-subject emotion recognition that relaxes the requirement for unlabeled data and learns subject-specific emotional representations from few labeled samples from the target subject.

\subsection{Inter-Subject Variance in EEG Emotion Recognition } 
Inter-subject variability in EEG emotion recognition arises mainly from individual differences in brain structure and cognitive processes. These variations pose significant challenges for developing generalized EEG emotion recognition models. Recently, domain adaptation (DA) and domain generalization (DG) techniques have emerged as promising solutions to mitigate the impact of inter-subject variance~\cite{sarafraz2024domain}. DA-based methods~\cite{9904850,10609541,10839595} aim to bridge the distributional gaps between source and target subjects to improve the performance of the target subject. They leverage various divergence metrics (e.g., MMD~\cite{borgwardt2006integrating}) to quantify and minimize distributional shifts. Chen \textit{et al.} introduced a multi-source marginal distribution strategy to preserve the marginal distributions of EEG data during adaptation~\cite{chen2021ms}. Li \textit{et al.} proposed a dynamic domain adaptation approach, which dynamically adapts domains based on class-aware information~\cite{9904850}. Wu \textit{et al.} developed the graph orthogonal purification network, which aligns the distribution of emotional features across dual spaces~\cite{10609541}. However, DA-based methods are limited in applicability due to their reliance on target subject information during training. Unlike DA, DG-based methods~\cite{10096469,10453943,10750375,liu2024moge,10856415} attempt to learn generalized models that perform well across multiple domains without requiring explicit adaptation to specific target domains. Chen \textit{et al.} proposed a graph domain disentanglement network to learn subject-invariant representations~\cite{10453943}. Ugan~\cite{10750375} was introduced to comprehensively model and constrain potential uncertain statistical shifts across individuals. Liu \textit{et al.} developed a sparse mixture of graph experts model, which learns transferable features by decomposing brain regions into simpler functional units~\cite{liu2024moge}. Despite their effectiveness, these methods remain in suboptimal performance due to the absence of target domain information. This fundamental limitation drives our exploration of few-shot learning paradigms for EEG-based emotion recognition, particularly in scenarios where only minimal labeled target subject data is available.
\subsection{Few-shot Learning}
Few-shot learning (FSL) is an advanced paradigm that facilitates rapid model adaptation to novel classes with only a few labeled examples~\cite{chen2021meta,wang2023improving,10445009}. This paradigm has recently gained attention in brain-computer interface research. Ning \textit{et al.} pioneered its application to cross-subject EEG-based emotion recognition by employing a prototypical network~\cite{snell2017prototypical} for efficient cross-subject generalization~\cite{ning2021cross}. Zhang \textit{et al.} developed EmoDSN~\cite{9751421}, a deep Siamese network architecture that achieves rapid convergence with minimal training data while maintaining robust performance in fine-grained emotion recognition tasks. These metric-learning-based methods, which focus on learning a metric space that generalizes to new subjects without requiring fine-tuning, exhibit similarities with DG-based methods~\cite{khoee2024domain}. Current research predominantly focuses on meta-learning strategies for cross-subject EEG emotion recognition. MetaEmotionNet~\cite{10342627} and MAML-EEG~\cite{Chen_2025} adapted the model-agnostic meta-learning~\cite{finn2017model} framework to EEG emotion recognition, demonstrating promising performance on novel subjects. FreAML~\cite{10822213} introduced a Bayesian meta-learning framework that derives task-specific model parameters from the frequency domain. SIML~\cite{ng2024subject} further hybridized conventional supervised learning with meta-learning to optimize EEG classifier training. These methods generally aim to learn a subject-agnostic initialization that can quickly adapt to new subjects with few labeled samples. However, adapting all model parameters can be sensitive to the variability of EEG signals under few-shot condition. To improve the stability and efficiency of subject adaptation under such constraints, this article adopts a partially updated adaptation strategy that optimizes only a compact subject-specific adapter for the target subject, while the shared backbone remains fixed. This allows the model to focus on fine-grained personalization without the overfitting risks associated with full-model updates.

\section{THE ADOPTED METHODOLOGY}
\label{method}
\begin{figure*} [t]
	\centering
	\includegraphics[scale=0.6]{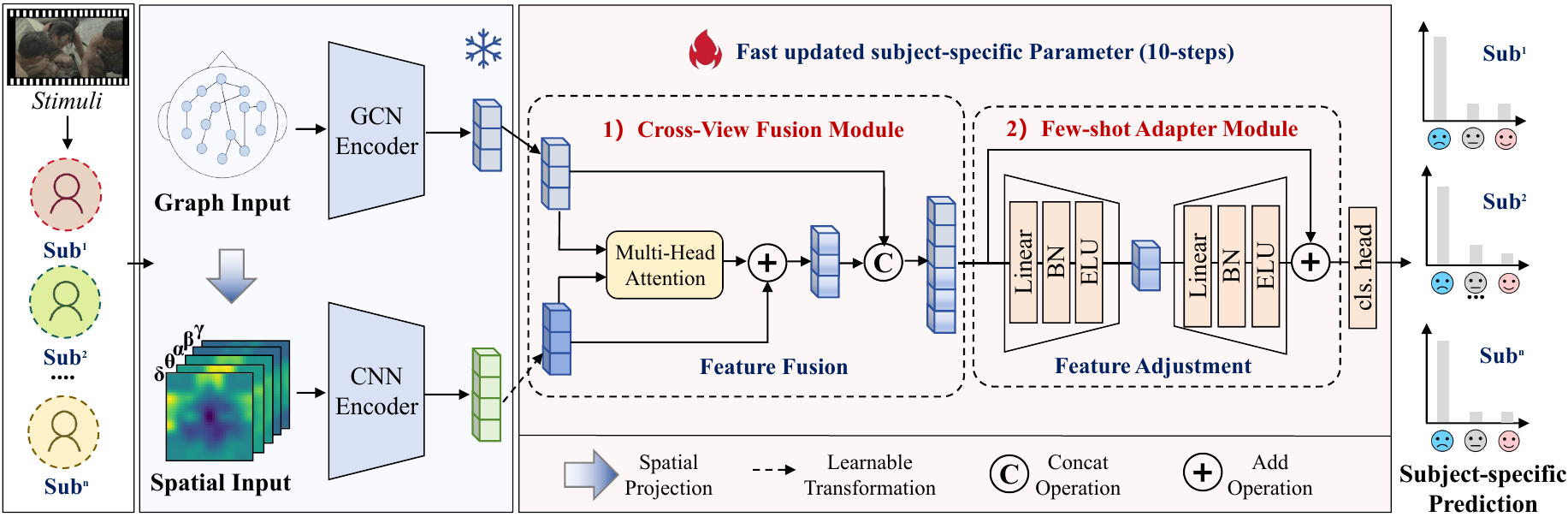}
	\caption{Overview of the proposed FACE architecture in the testing stage. For each unseen subject, a few labeled EEG signals are collected for rapid partial updates. Their EEG signals are spatially projected to spatial representations, which are then fed into FACE. The framework contains two core components: 1) CVF learns subject-specific fusion of global graph and local spatial features, and 2) The Few-Shot Adapter Module adjusts the features to ensure subject-specific emotional cues are captured.
	}
	\label{fig:framework}
\end{figure*}

This section introduces a few-shot adapter with a cross-view fusion approach for cross-subject EEG emotion recognition. The overall architecture of the proposed FACE framework is illustrated in Fig.~\ref{fig:framework}. FACE consists of two core components: the cross-view fusion and the few-shot adapter module. The following sections provide a detailed explanation of each component.

\subsection{Problem Definition}
In the cross-subject EEG emotion recognition task, each subject can be considered a distinct domain due to substantial inter-subject  variance. Accordingly, the data is divided into a source domain for training and a target domain for testing. Let the source domain be defined as $\mathcal{D}_s=\left\{\left(\mathbf{x}_i, y_i, S_i\right)\right\}_{i=1}^N,$ where $S_i \in S_{\text{source}}$ denotes the $i$-th source subject, $\mathbf{x}_i \in \mathcal{X}$ represents EEG signals, and $y_i \in \mathcal{Y}$ is the corresponding emotion label. The objective of cross-subject emotion recognition is to learn a model $f_\theta: \mathcal{X} \rightarrow \mathcal{Y}$ trained on $\mathcal{D}_s$ that generalizes to unseen target subjects $S_t \in S_{\text{target}}$, where ${S}_{\text {source}} \cap {S}_{\text {target }}=\emptyset$. The core challenge arises from domain shifts between source and target subjects, characterized by $P\left(\mathbf{X} \mid S_s\right) \neq P\left(\mathbf{X} \mid S_t\right),$ where $S_s \in S_{\text{source}}$. Existing methods often leverage auxiliary data from the target domain, such as a few labeled 
$\mathcal{D}_{t}^{l}=\left\{\left(\mathbf{x}_i, y_i, S_t\right),i=1,..,j\right\}$ and all unlabeled data $\mathcal{D}_{t}^{u}
=\left\{\mathbf{x}, S_t\right\}_{,}$ to mitigate distribution shifts. The few-shot cross-subject emotion recognition extends this task by imposing a strict few-shot constraint. For target subject $S_t$, the model must perform few-steps adaptation utilizing only a small support set $\mathcal{D}_{{t}}^{\text{sup}}=\left\{\left(\mathbf{x}_i, y_i\right)\right\}_{i=1}^{k \times c},$ where $k$ ($k \leq 10$) represents the number of labeled samples per emotion class $c$, and generalize to the query set $\mathcal{D}_{t}^{\text{que}}=\left\{\left(\mathbf{x}_i, y_i\right)\right\}_{i=1}^M.$ Here, $M$ represents the total number of remaining unlabeled samples from the target subject after excluding $\mathcal{D}_{{t}}^{\text{sup}}$. This paradigm is referred to as a $c$-way $k$-shot task. Formally, the model $f_\theta$ initially trained on the source domain $\mathcal{D}_s$ must generalize to target subject $S_t$ via few steps optimization using limited labeled samples $\mathcal{D}_{{t}}^{\text{sup}}$, such that:
\begin{equation}
	\min _{\theta^{\prime}} \mathbb{E}_{(\mathbf{x}, y) \sim \mathcal{D}_{t}^{\text {sup }}}\left[\mathcal{L}\left(f_{\theta^{\prime}}(\mathbf{x}), y\right)\right],
\end{equation}
where $\mathbb{E}$ denotes the empirical expectation. $f_{\theta^{\prime}}$ is the updated model and $\mathcal{L}(\cdot)$ is the loss function, i.e., cross-entropy loss.
\subsection{GCN Backbone}
Current methodologies for EEG-based emotion recognition commonly employ GCNs to capture discriminative brain topology from EEG signals. This article employs the dynamic GCN (DGCN)~\cite{9857970} as the main encoder. EEG signals are pre-processed into differential entropy (DE) features across multiple frequency bands, forming a representative feature $\mathbf{X} \in\mathbb{R}^{N\times C\times B}$ where $N$, $C$, and $B$ denote the number of samples, channels, and frequency bands, respectively. Each EEG sample $x \in \mathbf{X}$ can be represented as a graph $\mathcal{G}=(\mathcal{V}, \mathcal{E})$, where the node set $\mathcal{V}=\left\{v_1, \ldots, v_C\right\}$ corresponds to EEG  channels, and the edges $\mathcal{E}$ are weighted by a dynamic adjacency matrix $\mathbf{A}_d \in \mathbb{R}^{C \times C}$ that quantifies pairwise channel dependencies. The dynamic adjacency matrix can be formulated as:
\begin{equation}
	\mathbf{A}_d=\sigma_1\left(\mathbf{W}_2 \cdot \sigma_2\left(\mathbf{W}_1 \mathbf{A}_0\right)\right),
\end{equation}
where $\sigma_1$ and $\sigma_2$ denote different activation functions. $\mathbf{A}_0$ represents a randomly initialized matrix, and $\mathbf{W}_1 \in \mathbb{R}^{\left({C\over r} \times C\right)}$  and $\mathbf{W}_2 \in \mathbb{R}^{C \times \left(C\over r\right)}$ are learnable projection matrices with a reduction ratio $r$.

The dynamic GCN processes $\mathbf{X}$ through a two-layer GCN to derive high-level representations which can be formally expressed as:
\begin{equation}
	\mathbf{H}={\operatorname{BN}}\left(\boldsymbol{\Theta}_2 * \sigma \left( \operatorname{BN}\left(\boldsymbol{\Theta}_1 * \mathbf{X} \right)\right)\right),
\end{equation}
where $\boldsymbol{\Theta}_1$, $\boldsymbol{\Theta}_2 \in \mathbb{R}^{1\times K}$ are convolution kernels with stride 1, $*$ denotes convolution operation, and BN represents the batch normalization layer. 
The node aggregation is performed with residual operation to obtain the hierarchical feature representations, which can be expressed as:
\begin{equation}
	\label{eq:gcn_z}
	\mathbf{Z}_g = \operatorname{BN} \left(\mathbf{L}\mathbf{H}+\mathbf{X}\right),
\end{equation}
where $\mathbf{L} =\mathbf{D}^{-1}\mathbf{A}_d $ denotes the graph Laplacian. The hierarchical feature $\mathbf{Z}_g$ is finally fed into a fully connected layer for final emotion classification. The final emotion prediction $\widehat{y}$ is obtained via:
\begin{equation}
	\widehat{y} = \text{Softmax} \left( f_c(\mathbf{Z}_g) \right).
\end{equation}
And the model is optimized by minimizing the cross-entropy loss as follows:
\begin{equation}
	\mathcal{L}_{\mathrm{cls}}=-\frac{1}{N} \sum_{i=1}^N \sum_{j=1}^c y_{i, j} \log \left(\hat{y}_{i, j}\right),
\end{equation}
where $y_{i,j}$ is the ground-truth one-hot label of the $i$-th sample, and $\widehat{y}_{i,j}$ represents the prediction for class $j$.

\subsection{Cross-view Fusion Module}
While DGCN effectively models global brain topology, local spatial patterns provide complementary information for EEG emotion recognition. We therefore introduce a spatial branch and develop the CVF module to align and integrate the two views. Specifically, CVF uses the graph representation to refine the spatial features and then combines the refined spatial features with the original graph representation, thereby preserving their complementary information. Since the relationship between the two views varies across subjects, the fusion parameters are optimized through meta-learning and adapted to each unseen subject using a few labeled samples.

\subsubsection{Spatial Feature Extraction}
EEG signals are recorded from multiple scalp-mounted electrodes. Each electrode monitors electrical activity in specific brain regions. The spatial arrangement of these electrodes preserves the brain's intrinsic topographic organization. These spatial patterns in EEG data reveal critical insights into emotion-related functional connectivity and neural activation dynamics. It is natural to leverage such patterns to provide auxiliary information for more precise prediction.

Given the input EEG signals $\mathbf{X} \in \mathbb{R}^{N\times C \times B}$, this article first performs spatial projection $\Phi( \cdot)$ based on physical electrode arrangements~\cite{ning2021cross,10568943}:

\begin{equation}
	\Phi: \mathbb{R}^{N\times C \times B} \rightarrow \mathbb{R}^{N\times B \times H\times W}, \quad \mathbf{S} = \Phi(\mathbf{X}),
\end{equation}
where $\mathbf{S}$ represents the spatial form of $\mathbf{X}$. We apply a three-layer CNN encoder to obtain the deep spatial feature $\mathbf{Z}_s$ from $\mathbf{S}$. To preserve fine-grained spatial resolution, we omit the final pooling operation and flatten the spatial feature in the channel dimension. Finally, the deep spatial feature $\mathbf{Z}_s \in \mathbb{R}^{N \times D}$ is utilized for further fusion.

\subsubsection{Cross-view Fusion Mechanism}
The significant inherent intra-subject variance in EEG signals introduces a critical challenge for fusing heterogeneous views. Existing approaches typically process these views independently or employ simple fusion strategies, failing to capture their complex inter-dependency and potentially introducing feature misalignment. Moreover, inter-subject variability further complicates the fusion, as the relationship between the two views may shift across subjects. Drawing inspiration from dynamic fusion strategies~\cite{ye2024sg,10384690} in other fields, this article proposes a novel cross-view fusion mechanism that dynamically generates a unified representation from heterogeneous views for unseen subjects, thereby bridging this gap.

The global graph representation $\mathbf{Z}_g$ obtained from
	Equation~(\ref{eq:gcn_z}) and the deep spatial feature $\mathbf{Z}_s$
	characterize EEG signals from complementary views but have different feature
	dimensions. We therefore project $\mathbf{Z}_s$ to the same dimension as
	$\mathbf{Z}_g$ by learnable linear transformation:
\begin{equation}
	\tilde{\mathbf{Z}}_s = \mathbf{W}_{sg}\mathbf{Z}_s + b_{sg}.
\end{equation}

Dimension matching alone cannot eliminate the discrepancy between the two
views. We thus employ multi-head attention to capture the cross-view
dependencies between $\widetilde{\mathbf{Z}}_s$ and $\mathbf{Z}_g$. For the
$l$-th attention head, the relation score is calculated as
\begin{equation}
	A_{sg}^{(l)}
	=
	\frac{
		\exp\left(
		\left\langle
		Q_s^{(l)},K_g^{(l)}
		\right\rangle/\sqrt{d_h}
		\right)
	}{
		\exp\left(
		\left\langle
		Q_s^{(l)},K_s^{(l)}
		\right\rangle/\sqrt{d_h}
		\right)
		+
		\exp\left(
		\left\langle
		Q_s^{(l)},K_g^{(l)}
		\right\rangle/\sqrt{d_h}
		\right)
	}.
\end{equation}
Here, $Q_s^{(l)}$ and $K_s^{(l)}$ are obtained from
$\widetilde{\mathbf{Z}}_s$, while $K_g^{(l)}$ is obtained from
$\mathbf{Z}_g$. The attention weight $A_{sg}^{(l)}$ measures the dependency
between the spatial and graph views for each sample. The cross-view dependencies captured by different attention heads are
aggregated as
\begin{equation}
	\mathbf{Z}_{\triangle}
	=
	\mathbf{W}_o
	\left[
	A_{sg}^{(1)}\odot V_g^{(1)};
	\cdots;
	A_{sg}^{(L)}\odot V_g^{(L)}
	\right],
\end{equation}
where $V_g^{(l)}$ is the value projection of $\mathbf{Z}_g$, $L$ denotes
the number of attention heads, $d_h$ is the feature dimension of each head,
and $\mathbf{W}_o$ combines the head outputs. Consequently,
$\mathbf{Z}_{\triangle}$ encodes the dependencies between the global
connectivity and spatial views. We use these cross-view dependencies to refine the spatial representation:
\begin{equation}
	\widetilde{\mathbf{Z}}_s'
	=
	\widetilde{\mathbf{Z}}_s+\mathbf{Z}_{\triangle}.
\end{equation}

After cross-view alignment, the two representations are
concatenated to obtain the unified representation:
\begin{equation}
	\mathbf{Z}_u
	=
	[\widetilde{\mathbf{Z}}_s';\mathbf{Z}_g].
\end{equation}

To reduce the effect of inter-subject differences on cross-view fusion,
the parameters of the CVF module are optimized as shared meta-parameters
using model-agnostic meta-learning~\cite{finn2017model}. Through episodic
bi-level optimization, CVF learns transferable cross-view fusion that
can be adapted to a new subject using only a few labeled samples. The
detailed optimization procedure is presented in the following subsection.

\subsection{Few-shot Adapter Module}
Acquiring a subject-agnostic unified representation via the CVF module remains insufficient due to the pronounced inter-subject variability in EEG signals.  While unified representations provide comprehensive features, they necessitate further refinement to capture discriminative subject-specific emotional patterns. Inspired by adapter-based architectures~\cite{li2022cross} in CV, this work devises a novel few-shot adapter integrated with meta-learning principles to preserve individualized emotional features. 

This article enhances conventional residual adapters through the strategic integration of batch normalization layers to statistically regularize inter-subject variance as follows:
\begin{equation}
	\mathcal{A}(\mathbf{Z}_u) = \sigma(\operatorname{BN}(\mathbf{W}_{l2} \cdot \sigma (\operatorname{BN}(\mathbf{W}_{l1}\mathbf{Z}_u))))+ \mathbf{Z}_u,
	\label{eq:adapter}
\end{equation}
where $\mathbf{W}_{l1} \in \mathbb{R}^{d_a \times d_u}$ and  $\mathbf{W}_{l2} \in \mathbb{R}^{d_u \times d_a}$ $(d_a\ll d_u)$ enforce a bottleneck for efficient adaptation. The stacked batch normalization layers statistically regularize cross-subject distribution shifts by normalizing feature statistics. For clarity, we omit bias terms in Eq.(\ref{eq:adapter}). The predicted emotion can be computed using a fully connected layer:
\begin{equation}
	\widehat{y} = \text{Softmax}\!\left(f_c(\mathcal{A}(\mathbf{Z}_u))\right).
\end{equation}

\begin{figure} [!t]
	\centering
	\includegraphics[scale=0.55]{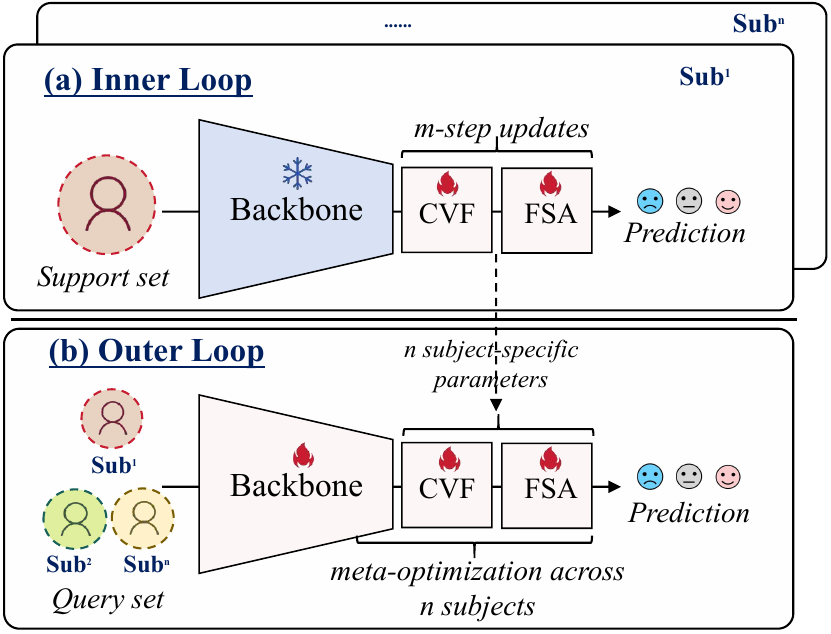}
	\caption{Illustration of the FACE training process: (a) inner-loop subject adaptation on the support set, and (b) outer-loop meta-update using query losses across $n$ subjects.
	}
	\label{fig:training}
\end{figure}
To enable the few-shot adapter to rapidly capture subject-specific emotional cues, we adopt a meta-learning strategy based on MAML~\cite{finn2017model}. We reformulate subject adaptation as a partial meta-learning task, which enables a subset of model components to be meta-optimized for rapid subject-specific adjustment. As illustrated in Fig.~\ref{fig:training}, the training follows a bi-level optimization scheme: an inner loop adapts the model to individual subjects, while an outer loop learns a subject-agnostic initialization. Under this formulation, each subject is considered as a separate task $\mathcal{T}_i \sim p(\mathcal{T})$.

We define the meta-optimized parameters as $\Theta = \{ \theta_\text{M}, \theta_\mathcal{A}, \theta_{f_c} \}$, which correspond to the multi-head attention, the few-shot adapter, and the classification head, respectively. The full model parameters are $\Theta_\text{all} = \{ \theta_{\text{gcn}}, \theta_{\text{cnn}}, \Theta\}$, where $\theta_{\text{gcn}}$ and $\theta_{\text{cnn}}$ are pre-trained weights from the GCN and CNN encoders. During each training epoch, we sample $n$ subjects. For each subject $\mathcal{T}_i$, we construct a support set $\mathcal{D}_i^{\text{sup}}$ and a query set $\mathcal{D}_i^{\text{que}}$.

(Inner-loop optimization): For every subject $\mathcal{T}_i$, the partial parameters $\Theta$ are adapted via $m$ gradient steps on $\mathcal{D}_i^{\text{sup}}$:
\begin{equation}
	\Theta'_i = \Theta - \alpha \nabla_\Theta \mathcal{L}_{\mathcal{T}_i}(f_{\Theta_\text{all}}(\mathcal{D}_i^\text{sup}), y),
\end{equation}
where $\alpha = 0.01$ is the inner-loop learning rate.

(Outer-loop optimization): After performing $n$ inner-loop adaptations, the meta-initialization $\Theta$ is updated by optimizing the aggregate performance of the adapted models ${ \Theta'_i }$ on their respective query sets $\mathcal{D}_i^{\text{que}}$. The goal is to learn an initialization that can rapidly converge to subject-specific optima with only a few gradient steps. To this end, the outer loop evaluates all inner-loop–adapted models on their query sets and integrates the resulting second-order gradient information across the $n$ subjects. This meta-optimization is formally expressed as:

\begin{equation}
	\min_{\Theta_\text{all}} \sum_{\mathcal{T}_i} \mathcal{L}_{\mathcal{T}_i}(f_{\Theta'_\text{all\_i}}(\mathcal{D}_i^\text{que}), y).
\end{equation}
Here, $\Theta'_{all\_i} = \{\theta_{\text{gcn}}, \theta_{\text{cnn}}, \Theta'_i \}$. Notably, the labels of $\mathcal{D}_i^{\text{que}}$ are available for the source subjects. The meta-optimization updates whole model parameters through second-order gradients:

\begin{equation}
	\Theta'_\text{all} \leftarrow \Theta_\text{all} - \beta \nabla_{\Theta_\text{all}} \sum_{\mathcal{T}_i} \mathcal{L}_{\mathcal{T}_i}(f_{\Theta'_\text{all\_i}}(\mathcal{D}_i^\text{que}), y),
\end{equation}
with $\beta=0.001$ denoting the outer loop learning rate. The proposed framework yields a subject-agnostic parameter initialization  $\Theta'_\text{all}$, via $N_e$ episodes of bi-level optimization during the meta-training phase. The complete training procedure is formalized in Algorithm~\ref{alg:meta_training}.

During cross-subject evaluation, the proposed framework demonstrates efficient adaptation through its meta-learned parameter initialization $\Theta'_\text{all}$. When encountering novel target subjects $S_t$, only $k$-shot labeled EEG samples per class need to be collected as the support set $\mathcal{D}_{t}^\text{sup}$ to perform rapid parameter adaptation via inner-loop optimization:

\begin{equation}
	\Theta'_t = \Theta - \alpha \nabla_\Theta \mathcal{L}_(f_{\Theta'_\text{all}}(\mathcal{D}_t^\text{sup}), y).
\end{equation}
When $k=3$, the proposed framework achieves performance comparable to state-of-the-art methods.

\begin{algorithm}[H]
	\caption{Training Procedure of the Proposed Method}
	\label{alg:meta_training}
	\begin{algorithmic}[1]
		\State Initialize $\Theta_{all} = \{ \theta_{\text{gcn}}, \theta_{\text{cnn}}, \theta_\text{M}, \theta_{\mathcal{A}}, \theta_{f_c} \}$
		\State Set inner-loop and outer-loop learning rate $\alpha$, $\beta$
		\For{episode $e = 1$ to $N_e$}   \Comment{Outer-loop}
		\State Sample $n$ subjects $\{ \mathcal{T}_1, \mathcal{T}_2, \cdots, \mathcal{T}_n \}$ from source subject distribution $p(\mathcal{T})$
		\For{$i = 1$ to $n$}
		\State Sample support set $\mathcal{D}_i^\text{sup}$ and query set $\mathcal{D}_i^\text{que}$ for subject $\mathcal{T}_i$
		\State Let $\Theta = \{ \theta_\text{M}, \theta_{\mathcal{A}}, \theta_{f_c} \}$
		\For{$j = 1$ to $m$} \Comment{Inner-loop}
		\State $\Theta'_i = \Theta - \alpha \nabla_\Theta \mathcal{L}_{\mathcal{T}_i}(f_{\Theta_\text{all}}(\mathcal{D}_i^\text{sup}), y)$
		\EndFor
		\State Store $\Theta'_\text{all\_i} = \{ \theta_\text{gcn}, \theta_\text{cnn}, \Theta'_i \}$
		\EndFor
		\State Compute meta-gradient:
		\State \hspace{1em} $g \leftarrow \nabla_{\Theta_\text{all}} \sum_{\mathcal{T}i} \mathcal{L}_{\mathcal{T}_i}(f_{\Theta'_\text{all\_i}}(\mathcal{D}_i^\text{que}), y)$
		\State Update all parameters: $\Theta'_\text{all} \gets \Theta_\text{all} - \beta g$
		\EndFor
		\State \Return $\Theta'_\text{all}$
	\end{algorithmic}
\end{algorithm}

\section{PERFORMANCE EVALUATION}
\label{exp_result}
The benchmarks are described in Section~\ref{sec:dataset}, followed by a detailed implementation of the proposed approach in Section~\ref{sec:setup}. Comparative experimental results with different methods are presented in Sections~\ref{result_seed_seediv} ,~\ref{result_seed5} and~\ref{result_ttest_methods}, respectively.

\subsection{Datasets}
\label{sec:dataset}
This article evaluates the performance of the proposed approach across three well-established EEG emotion recognition benchmarks, including SEED~\cite{7104132}, SEED-IV~\cite{8283814}, and SEED-V~\cite{liu2021comparing}. Table~\ref{tb:dataset} summarizes the key statistics of the datasets used in this study. A detailed description of these benchmarks is provided below:
\begin{itemize}
	\item {\bf SEED}~\cite{7104132}: The SEED dataset is an EEG emotion recognition dataset collected using the ESI NeuroScan System with a 62-channel electrode cap. It includes data from 15 participants (7 males, 8 females), each participating in 3 sessions approximately one week apart. During each session, participants watched 15 film clips with 3 different emotional tendencies (negative, positive, and neutral), for a total of 45 trials per participant. Following the previous setup, this study employs preprocessed DE features and segments the data into 1-second, non-overlapping windows.
	\item {\bf SEED-IV}~\cite{8283814}: The SEED-IV dataset extends SEED, capturing EEG signals from videos with 4 emotional tendencies: happiness, sadness, fear, and neutrality. It includes data from 15 participants (7 males, 8 females), each participating in 3 sessions. Consistent with previous work, this study utilizes preprocessed DE features and segments each session into 4-second, non-overlapping windows for analysis.
	\item {\bf SEED-V}~\cite{liu2021comparing}: The SEED-V dataset extends SEED-IV, with EEG data from 16 participants (6 males, 10 females) over 3 sessions. The emotional categories increase to 5 (happiness, sadness, disgust, neutrality, and fear). Consistent with previous setups, this study utilizes preprocessed DE features and segments the data into 4-second, non-overlapping windows for analysis.
\end{itemize}
\begin{table}[!htbp]
	\caption{Statistics of the datasets used in this study.}
	\centering
	\renewcommand{\arraystretch}{1.5}
	\setlength{\tabcolsep}{2.5pt}
	\label{tb:dataset}
	\begin{tabular}{ccccc}
		\toprule
		\textbf{Dataset} &
		\makecell{\textbf{Num of}\\\textbf{Subjects}} &
		\makecell{\textbf{Data per Subject}\\\textbf{per Session}} &
		\makecell{\textbf{Official}\\\textbf{Length}} &
		\makecell{\textbf{DE Samples per}\\\textbf{Subject per Session}} \\
		\hline
		SEED    & 15 & 5 trials × 3 classes & 1s & 3394      \\
		SEED-IV & 15 & 6 trials × 4 classes & 4s & 822$\sim$851   \\
		SEED-V  & 16 & 3 trials × 5 classes & 4s & 541$\sim$681   \\
		\bottomrule
	\end{tabular}
\end{table}

\subsection{Experiment Setup}
\label{sec:setup}
\textbf{Network Architecture:} This article employs DGCN and Conv3 as the main backbone. Following previous work~\cite{9857970}, the DGCN encoder utilizes a two-layer residual architecture without incorporating $K$-order Chebyshev polynomials. For this branch, DE features from five frequency bands are concatenated to form the input of size $N\times62\times5$, where $N$ is the number of samples. Conv3 consists of 3 blocks. Each block contains a $3\times3$ kernel, a batch normalization layer, a ReLU activation layer, and a $2\times2$ max-pooling layer. To enable effective spatial learning from EEG signals, the 62-channel DE feature is first projected onto a 2D grid following the physical electrode arrangement on a $9\times9$ map~\cite{ning2021cross}. This spatial representation is subsequently interpolated to $32\times32$ resolution. The final input size of the Conv3 branch is $N\times5\times32\times32$.

\textbf{Evaluation Protocol:} 
This article conducted cross-subject experiments to evaluate the generalization performance of FACE. Following previous studies, this article employed a leave-one-subject-out (LOSO) strategy. In LOSO, each subject serves as the target for testing, while the remaining subjects serve as the training set. The final performance is reported as the average across all target subjects. Given the few-shot learning paradigm, the model requires a few labeled samples from the target subject. This article randomly shuffles the target subject’s samples and selects $K$ labeled samples per class for adaptation, with the remaining samples used for testing. Despite using extremely limited labeled samples (approximately 1\%/5\%/8\% of the SEED/SEED-IV/SEED-V datasets under the 10-shot setting), the proposed method achieves performance comparable to representative domain adaptation and domain generalization approaches. Meanwhile, we also conduct a fair comparison with state-of-the-art few-shot learning methods under the same experimental protocol, fully demonstrating the effectiveness of our approach in the few-shot setting. To mitigate the risk of overfitting, this article repeated the experiment 200 times for each subject and reported the average performance.

\textbf{Implementation Details:} 
Recent work~\cite{chen2021meta,ng2024subject} in few-shot learning commonly adopts a supervised pre-training to provide a stable initialization before meta-training. This article utilizes a two-stage approach involving pre-training and few-shot learning. In the pre-training stage, the backbone is trained in a conventional supervised manner on whole-subject emotion classification using smooth cross-entropy loss~\cite{muller2019does}. We employ the Adam optimizer with a learning rate $0.001$ and train the model from scratch for $50$ epochs. After pre-training, the classification head is removed and the pretrained encoders are used as the backbone initialization for the subsequent few-shot learning phase. In the few-shot learning stage, the model is further optimized using the Adam optimizer over $50$ episodes. For each episode, we randomly sample $N$ subjects. From each subject, $(K+Q)$ samples per class are randomly selected, where $K$ samples are used for a $10$-step fast update, and $Q$ samples are used to evaluate the updated model. The smooth cross-entropy loss is computed across the $N$ subjects to optimize the model, aiming to learn a good initialization that generalizes across all source subjects. In our experiments, $N$ is optimally set to $2$, $K$ is chosen from $\left \{1,3,5,10\right\}$, and $Q$ is set to $20$ during training and to the remaining samples of the target subject during testing. All experiments are implemented using the PyTorch framework and conducted on an AMD Instinct MI250 GPU.

\begin{table*}[!htbp]
	\caption{Comparison with state-of-the-art approaches on SEED and SEED-IV. The best results are shown in bold, and the second-best results are underlined.}
	\renewcommand{\arraystretch}{1.5}
	\centering
	\label{table_sota}
	\begin{tabularx}{\textwidth}{|>{\raggedright\arraybackslash}p{2.0cm}|X|>{\centering\arraybackslash}p{2.8cm}|>{\centering\arraybackslash}p{2.8cm}|>{\centering\arraybackslash}p{2.2cm}|>{\centering\arraybackslash}p{2.2cm}} 
		\toprule
		\multicolumn{2}{c|}{\multirow{2}{*}{\textbf{Method}}} & \multirow{2}{*}{\textbf{\begin{tabular}[c]{@{}c@{}}\#Num of unlabeled \\ target samples\end{tabular}}} & \multirow{2}{*}{\textbf{\begin{tabular}[c]{@{}c@{}}\#Num of labeled \\ target samples\end{tabular}}} &  \multirow{2}{*}{\textbf{\begin{tabular}[c]{@{}c@{}}SEED \\ Mean / STD (\%)\end{tabular}}}&  \multirow{2}{*}{\textbf{\begin{tabular}[c]{@{}c@{}}SEED-IV\\ Mean / STD (\%)\end{tabular}}} \\
		\multicolumn{2}{c|}{} &  &  &  &  \\ \hline
		\multicolumn{1}{c|}{\multirow{3}{*}{Supervised Learning}} &\centering IAG~\cite{song2020instance} & \multirow{3}{*}{\XSolidBrush} & \multirow{3}{*}{\XSolidBrush} & 86.30 / 6.91 & 62.64 / 10.25 \\ \cline{2-2} \cline{5-6} 
		\multicolumn{1}{c|}{} & \centering GMSS~\cite{9765326} &  &  & 86.5 / 6.22 & 73.48 / 7.41 \\ \cline{2-2} \cline{5-6} 
		\multicolumn{1}{c|}{} &\centering  PGCN~\cite{10496191} &  &  & 84.59 / 8.68 & 73.69 / 7.16 \\ \hline
		\multicolumn{1}{c|}{\multirow{3}{*}{Semi-Supervised learning}} &\centering SDDA~\cite{9904850} & \multirow{3}{*}{ALL} & $\sim$225/$\sim$35 & 91.08 / 7.70 & 81.58 / 8.72 \\ \cline{2-2} \cline{4-6} 
		\multicolumn{1}{c|}{} &\centering SSML~\cite{li2023novel} &  & 10-shot & 88.59 & - \\ \cline{2-2} \cline{4-6} 
		\multicolumn{1}{c|}{} &\centering  FSA-TSP~\cite{10214058} &  & 3-shot & 93.55 / 5.03 & 87.96 / 5.18 \\ \hline
		\multicolumn{1}{c|}{\multirow{5}{*}{Domain Adaptation}} &\centering MS-MDA~\cite{chen2021ms} & \multirow{5}{*}{ALL} & \multirow{5}{*}{\XSolidBrush} & 89.63 / 6.79 & 59.34 / 5.48 \\ \cline{2-2} \cline{5-6} 
		\multicolumn{1}{c|}{} &\centering  UDDA$^\ast$~\cite{9904850} &  &  & 88.10 / 6.54 & 73.14 / 9.43 \\ \cline{2-2} \cline{5-6} 
		\multicolumn{1}{c|}{} &\centering SSAA-MSDA~\cite{10509712} &  &  & 90.11 / 7.32 & 76.23 / 9.02 \\ \cline{2-2} \cline{5-6} 
		\multicolumn{1}{c|}{} &\centering  PR-PL$^\ast$~\cite{10160130} &  &  & 93.66 / 4.48 & 74.46 / 4.56 \\ \cline{2-2} \cline{5-6} 
		\multicolumn{1}{c|}{} &\centering Grop~\cite{10609541} &  &  & 91.58 / 4.02 & 75.63 / 9.20 \\ \hline
		\multicolumn{1}{c|}{\multirow{4}{*}{Domain Generalization}} &\centering  GMoE~\cite{li2023sparse} & \multirow{4}{*}{\XSolidBrush} & \multirow{4}{*}{\XSolidBrush} & 84.60 / 9.30 & 71.20 / 8.50 \\ \cline{2-2} \cline{5-6} 
		\multicolumn{1}{c|}{} &\centering MOGE~\cite{liu2024moge} &  &  & 88.00 / 4.50 & 74.30 / 6.10 \\ \cline{2-2} \cline{5-6} 
		\multicolumn{1}{c|}{} &\centering GDDN~\cite{10453943} &  &  & 92.54 / 3.65 & 75.65 / 5.47 \\ \cline{2-2} \cline{5-6} 
		\multicolumn{1}{c|}{} &\centering  Ugan$^\ast$~\cite{10750375} &  &  & 93.11 / 4.13 & 77.16 / 4.73 \\ \hline
		\multicolumn{1}{c|}{\multirow{7}{*}{Few-shot Learning}} & \centering MetaBaseline$^\ast$~\cite{chen2021meta} & \multirow{7}{*}{\XSolidBrush} & 5-shot & 84.39 / 8.18 & 81.38 / 4.96 \\ \cline{2-2} \cline{4-6} 
		\multicolumn{1}{c|}{} &\centering MMR$^\ast$~\cite{wang2023improving} &  & 5-shot & 84.69 / 7.80 & 83.89 / 4.04 \\ \cline{2-2} \cline{4-6} 
		\multicolumn{1}{c|}{} &\centering MAML-EEG$^\ast$~\cite{Chen_2025} &  & 5-shot & 79.54 / 8.70 & 84.95 / 3.94 \\ \cline{2-2} \cline{4-6} 
		\multicolumn{1}{c|}{} &\centering SIML$^\ast$~\cite{ng2024subject}& & 5-shot &88.13 / 5.63   & 85.70 / 3.48 \\ \cline{2-2} \cline{4-6} 
		\multicolumn{1}{c|}{} &\centering  \multirow{3}{*}{\textbf{FACE (Ours)}} &  & 3-shot & 91.66 / 3.72 & 83.91 / 3.87 \\ \cline{4-6} 
		\multicolumn{1}{c|}{} &  &  & 5-shot & \underline{93.96 / 2.70} & \underline{ 89.51 / 3.13} \\ \cline{4-6} 
		\multicolumn{1}{c|}{} &  &  & 10-shot & \textbf{96.72 / 1.81} & \textbf{95.95 / 1.47} \\ \hline
		\multicolumn{6}{l}{\footnotesize $\ast$ denotes the results reproduced through our own implementation based on open-source code.} \\
		\multicolumn{6}{l}{\footnotesize \makecell[l]{$\sim$ indicates an approximation. Since sample amounts vary across subjects, SDDA uses half of the target labeled samples for semi-supervised learning, \\allowing only an estimation of labeled target samples.}}
	\end{tabularx}
\end{table*}

\subsection{Comparison With State-of-the-Art Approaches}
\label{sec:sota}
This study presents a comprehensive comparison between the proposed method and existing state-of-the-art approaches, including three supervised learning methods (IAG~\cite{song2020instance}, GMSS~\cite{9765326}, and PGCN~\cite{10496191}), three semi-supervised learning methods (SDDA~\cite{9904850}, SSML~\cite{li2023novel}, and FSA-TSP~\cite{10214058}), five DA-based methods (MS-MDA~\cite{chen2021ms}, UDDA~\cite{9904850}, SSAA-MSDA~\cite{10509712}, PR-PL~\cite{10160130}, and Grop~\cite{10609541}), four DG-based methods (GMoE~\cite{li2023sparse}, MOGE~\cite{liu2024moge}, GDDN~\cite{10453943}, and Ugan~\cite{10750375}) and four few-shot learning methods (Meta-baseline~\cite{chen2021meta}, MMR~\cite{wang2023improving}, MAML-EEG~\cite{Chen_2025}, and SIML~\cite{ng2024subject}). Since DA and DG-based methods adopt different assumptions about target data from FSL, their results are included as references rather than direct comparisons. All experiments were conducted on three widely adopted EEG-based emotion recognition benchmarks. This article reproduces some methods based on open-source code. To ensure fair comparisons and maintain consistency in experimental settings, DGCN was uniformly adopted as the backbone for the few-shot learning methods. Since 5-shot is a widely adopted setting in few-shot learning, most experiments in this article use the 5-shot model as the base. The performance evaluations were carried out on the SEED and SEED-IV benchmarks, as outlined in Section~\ref{result_seed_seediv}, and on the SEED-V benchmark, as detailed in Section~\ref{result_seed5}.

\subsubsection{Results on SEED and SEED-IV}
\label{result_seed_seediv}
Table~\ref{table_sota} presents a comprehensive comparison of various methods on SEED and SEED-IV datasets. Our method demonstrates competitive performance with state-of-the-art methods. Specifically, FACE achieves an impressive accuracy of 96.72\% on SEED and 95.95\% on SEED-IV under the 10-shot setting. Furthermore, the proposed method consistently outperforms all competing approaches in the 5-shot setting, proving the effectiveness of the few-shot learning paradigm and the superiority of the proposed framework. However, the semi-supervised method FSA-TSP performs slightly better than our FACE method in the 3-shot setting. This can be attributed to the ability of semi-supervised methods, such as FSA-TSP, to utilize both labeled and large amounts of unlabeled samples during training, thereby improving their performance in low-data scenarios. When the number of labeled samples increases to 5-shot, this performance gap is eliminated. These results demonstrate that while semi-supervised methods benefit from unlabeled data, our approach achieves exceptional performance even with limited labeled data,  highlighting the effectiveness of the proposed method in few-shot learning scenarios and its strong generalization capability.

\subsubsection{Results on SEED-V}
Since the SEED-V dataset was released only recently, there are relatively few comparative methods available, and the dataset itself is relatively small in scale. Therefore, this article conducts a separate comparative analysis on this dataset and includes additional experiments under the 1-shot setting. Table~\ref{seed5_sota} demonstrates the effectiveness of the proposed method on the SEED-V dataset. Under the 1-shot setting, the proposed method already achieves competitive performance. Furthermore, in the 3-shot setting, the proposed approach outperforms all others. In the 10-shot setting, the proposed method achieves 98.95\% accuracy using only approximately 8\% of the labeled target domain samples. It is also worth noting that the particularly strong performance on SEED-V aligns with the intrinsic characteristics of the dataset summarized in Table~\ref{tb:dataset}. Compared with SEED, SEED-V contains fewer trials per class and exhibits lower intra-class variability, and its longer official 4-second segments generally yield more stable DE features. These properties lead to clearer class boundaries and reduced sample-level variability, which collectively make the few-shot classification inherently easier than in SEED. The influence of segment length on feature stability and model performance is discussed in more detail in Section~\ref{sec:seg_len}. Overall, these results consistently highlight the superiority and strong generalization of the proposed method in few-shot learning scenarios.

\label{result_seed5}
\begin{table}[]
	\caption{Comparison with state-of-the-art approaches on SEED-V. The best results are shown in bold, and the second-best results are underlined.}
	\centering
	\label{seed5_sota}
	\renewcommand{\arraystretch}{1.5}
	\begin{tabular}{c|c|c|c}
		\hline
		\multirow{2}{*}{\textbf{Method}} & \multirow{2}{*}{\textbf{\#NUS}} & \multirow{2}{*}{\textbf{\#NLS}} & \multirow{2}{*}{\textbf{\begin{tabular}[c]{@{}c@{}}SEED-V \\ Mean / STD (\%)\end{tabular}}} \\
		&  &  &  \\ \hline
		IAG~\cite{song2020instance} & \multirow{3}{*}{\XSolidBrush} & \multirow{3}{*}{\XSolidBrush} & 59.68 / 9.44 \\ \cline{1-1} \cline{4-4} 
		Progressive GCN~\cite{zhou2023progressive} &  &  & 71.40 / 9.43 \\ \cline{1-1} \cline{4-4} 
		PGCN$^\ast$~\cite{10496191}  &  &  &65.21 / 10.10\\ \hline
		
		JAGP~\cite{9817639} & \multirow{3}{*}{ALL} & \multirow{3}{*}{\XSolidBrush} & 75.43 / 7.57 \\ \cline{1-1} \cline{4-4} 
		SSAA-MSDA$^\ast$~\cite{10509712} &  &  & 63.09 / 10.01 \\ \cline{1-1} \cline{4-4} 
		PR-PL$^\ast$~\cite{10160130} &  &  & 72.62 / 12.67 \\ \hline
		GMoE~\cite{li2023sparse} & \multirow{3}{*}{\XSolidBrush} & \multirow{3}{*}{\XSolidBrush} & 76.30 / 9.20 \\ \cline{1-1} \cline{4-4} 
		Ugan$^\ast$~\cite{10750375} &  &  & 77.12 / 10.87 \\ \cline{1-1} \cline{4-4} 
		MoGE~\cite{liu2024moge} &  &  & 81.80 / 10.00 \\ \hline
		MetaBaseline$^\ast$~\cite{chen2021meta} & \multirow{3}{*}{\XSolidBrush} & 5-shot & 85.16 / 5.64 \\ \cline{1-1} \cline{3-4} 
		MMR$^\ast$~\cite{wang2023improving} &  & 5-shot & 82.85 / 5.10 \\ \cline{1-1} \cline{3-4} 
		SIML$^\ast$~\cite{ng2024subject} &  & 5-shot & 84.88 / 4.59 \\ \hline
		\multirow{4}{*}{\textbf{FACE (Ours)}} & \multirow{4}{*}{\XSolidBrush} & 1-shot & 75.74 / 7.00 \\ \cline{3-4} 
		&  & 3-shot & 89.55 / 2.98 \\ \cline{3-4} 
		&  & 5-shot & \underline{95.20 / 1.42} \\ \cline{3-4} 
		&  & 10-shot & \textbf{98.95 / 0.36} \\ \hline
		\multicolumn{4}{l}{\footnotesize \makecell[l]{$\ast$ denotes the results reproduced through our own implementation based \\ on open-source code.}} \\
		\multicolumn{4}{l}{\footnotesize \makecell[l]{\textbf{\#NUS} refers to the number of unlabeled target samples and \textbf{\#NLS} refers \\ to the number of labeled target samples used for training.}}
		
	\end{tabular}
\end{table}

\subsubsection{Significance Test over Existing Methods}
\label{result_ttest_methods}
\begin{figure} [!htb]
	\centering
	\includegraphics[scale=0.36]{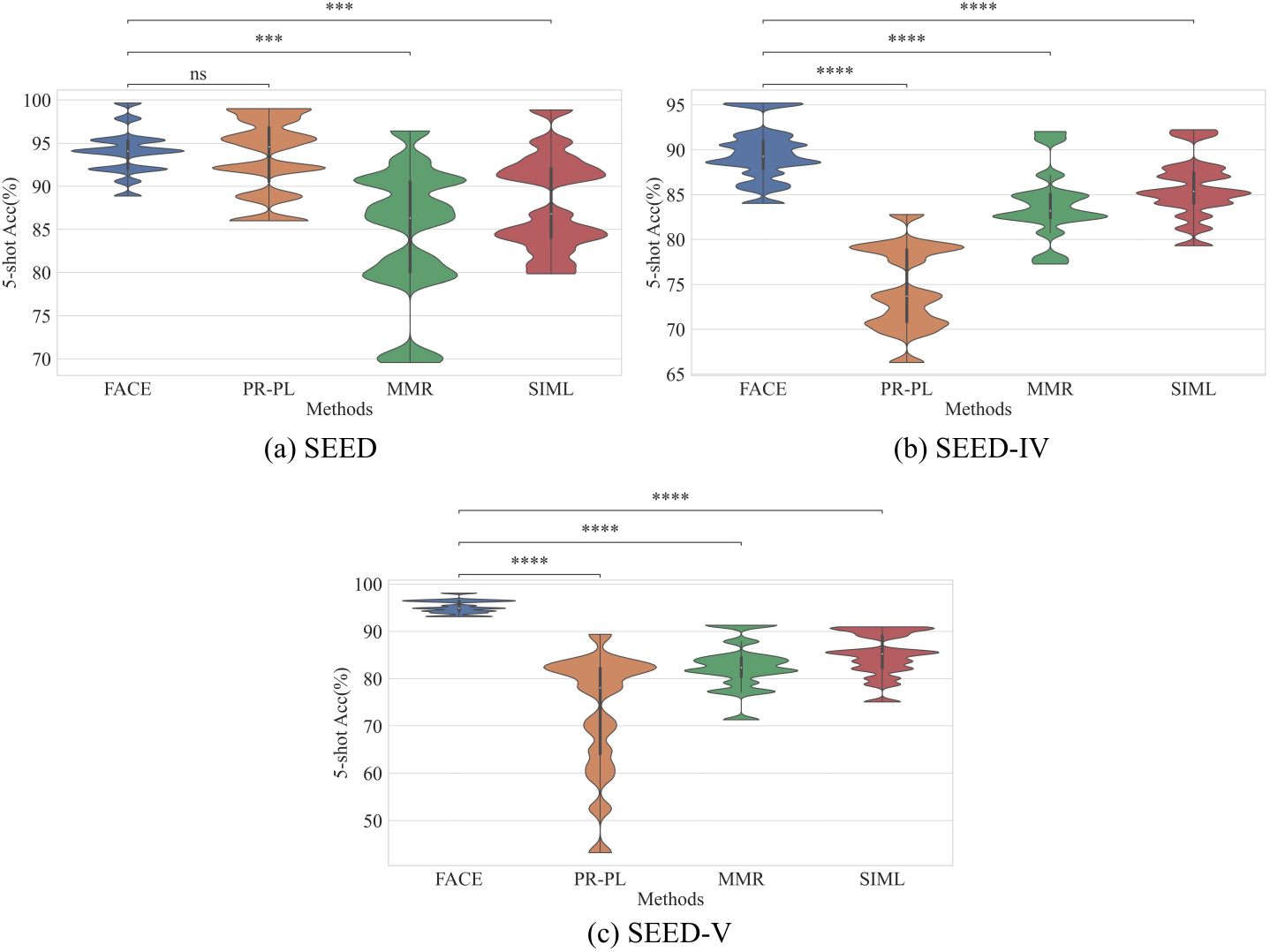}
	\caption{Statistical comparison of different methods on the (a) SEED, (b) SEED-IV, and (c) SEED-V datasets. The asterisks (***) indicate $0.0001 < p \leq 0.001$, and four asterisks (****) denote $p \leq 0.0001$, with more asterisks indicating stronger statistical significance. “ns” indicates no significant difference.
	}
	\label{fig:t-test—methods}
\end{figure}

Fig.~\ref{fig:t-test—methods} presents a significance test between the proposed FACE and three representative baselines across SEED, SEED-IV, and SEED-V. Beyond differences in accuracy, the significance patterns clearly reveal why FACE achieves consistent superiority. Across all three datasets, FACE not only achieves the highest median accuracy but also maintains tightly concentrated score distributions, leading to significant improvements over existing methods. This statistical advantage stems from FACE's subject-adaptive design, which performs fine-grained calibration based on limited support samples, thereby reducing sensitivity to cross-subject variability. An interesting observation is that the gap between FACE and PR–PL is non-significant on SEED but becomes significant on SEED-IV and SEED-V. This is largely due to SEED providing substantially more samples per subject, giving PR–PL access to richer unlabeled target data for domain adaptation and yielding more stable prototype alignment. In contrast, the smaller scale of SEED-IV and SEED-V limits such alignment, causing PR–PL to fluctuate and making FACE’s advantage statistically significant. Overall, the significance tests confirm that FACE not only improves average performance but also provides more stable few-shot adaptation, outperforming existing methods in both accuracy and consistency.

\section{Discussion}
\label{discussion}
This section presents a series of ablation studies and visualizations to evaluate the contribution of each module. Section~\ref{sec:module-wise} reports the module-wise ablation experiments, while Sections~\ref{sec:cvf} and~\ref{sec:fsa} further analyze the effects of the CVF and FSA modules, respectively. Sections~\ref{sec:dqd}, ~\ref{sec:num_shot}, and ~\ref{sec:seg_len} investigate the impact of data scale and diversity, the number of shots, and temporal resolution. Finally, Section~\ref{sec:Para} analyzes the influence of the hyperparameter $L$, which corresponds to the number of attention heads.

\subsection{Module-wise Ablation Study}
\label{sec:module-wise}
Table~\ref{ablation} presents the module-wise ablation studies of the proposed method on the SEED, SEED-IV, and SEED-V datasets. Across different shot settings, incorporating the CVF or FSA module alone generally improves performance compared to the baseline. This demonstrates that subject-specific optimization, whether applied to feature fusion or logit prediction, benefits the model performance. Optimal results are achieved when both the CVF and FSA modules are employed. The main reason is that the CVF module provides comprehensive subject-specific features, facilitating more effective adaptation by the FSA module. In this way, the model can learn subject-specific emotional cues more efficiently, thereby improving overall generalization performance. 
\begin{table*}[!htbp]
	\caption{Ablation studies of the proposed method on the SEED, SEED-IV, and SEED-V datasets, with the best results highlighted in bold.}
	\centering
	\label{ablation}
	\renewcommand{\arraystretch}{1.5}
	\begin{tabular}{ccccccccccc}
		\toprule
		\multicolumn{1}{c}{\multirow{2}{*}{CVF}} & \multicolumn{1}{c}{\multirow{2}{*}{FSA}} & \multicolumn{3}{c}{SEED} & \multicolumn{3}{c}{SEED-IV} & \multicolumn{3}{c}{SEED-V} \\ \cmidrule(lr){3-5}  \cmidrule(lr){6-8}  \cmidrule(lr){9-11}  
		\multicolumn{1}{c}{} & \multicolumn{1}{c}{} & \multicolumn{1}{c}{3-shot} & \multicolumn{1}{c}{5-shot} & \multicolumn{1}{c}{10-shot} & \multicolumn{1}{c}{3-shot} & \multicolumn{1}{c}{5-shot} & \multicolumn{1}{c}{10-shot} & \multicolumn{1}{c}{3-shot} & \multicolumn{1}{c}{5-shot} & \multicolumn{1}{c}{10-shot} \\ 
		\midrule 
		&  & \multicolumn{1}{c}{81.80{\scriptsize $\pm$7.79}}  & \multicolumn{1}{c} {86.60{\scriptsize  $\pm$6.27}}  & {89.70{\scriptsize  $\pm$5.36}}  & \multicolumn{1}{c} {80.37{\scriptsize  $\pm$4.33}}  & \multicolumn{1}{c} {84.65{\scriptsize  $\pm$3.75}}  & {88.21{\scriptsize  $\pm$3.07}}& \multicolumn{1}{c}{78.31{\scriptsize  $\pm$5.23}} & \multicolumn{1}{c}{82.63{\scriptsize  $\pm$4.70}} & {85.20{\scriptsize  $\pm$5.17}} \\
		\checkmark &  & \multicolumn{1}{c}{90.84{\scriptsize  $\pm$5.26}} & \multicolumn{1}{c}{93.70{\scriptsize  $\pm$3.85}} & \multicolumn{1}{c}{95.36{\scriptsize  $\pm$3.03}} & \multicolumn{1}{c}{82.57{\scriptsize  $\pm$4.44}} & \multicolumn{1}{c}{87.03{\scriptsize  $\pm$3.86}} & \multicolumn{1}{c}{90.73{\scriptsize  $\pm$3.26}} & \multicolumn{1}{c}{87.60{\scriptsize  $\pm$3.80}} & \multicolumn{1}{c}{91.08{\scriptsize  $\pm$3.38}} & \multicolumn{1}{c}{94.63{\scriptsize  $\pm$2.23}} \\ 
		&  \checkmark & \multicolumn{1}{c} {84.51{\scriptsize  $\pm$7.04}} & \multicolumn{1}{c} {88.86{\scriptsize  $\pm$5.48}} & {94.02{\scriptsize  $\pm$3.77}} & \multicolumn{1}{c}{83.48{\scriptsize  $\pm$4.16}} & \multicolumn{1}{c}{88.75{\scriptsize  $\pm$2.74}} & {95.10{\scriptsize  $\pm$1.30}} & \multicolumn{1}{c}{87.81{\scriptsize  $\pm$2.81}} & \multicolumn{1}{c}{94.09{\scriptsize  $\pm$1.58}} &{98.23{\scriptsize  $\pm$0.87}} \\
		\checkmark &  \checkmark & \multicolumn{1}{c}{\textbf{91.66{\scriptsize  $\pm$3.72}}} & \multicolumn{1}{c}{\textbf{93.96{\scriptsize  $\pm$2.70}}} & \textbf{{96.72{\scriptsize  $\pm$1.81}}} & \multicolumn{1}{c}{\textbf{83.91{\scriptsize  $\pm$3.87}}} & \multicolumn{1}{c}{\textbf{89.51{\scriptsize  $\pm$3.13}}} & {\textbf{95.95{\scriptsize  $\pm$1.47}}} & \multicolumn{1}{c} {\textbf{89.55{\scriptsize  $\pm$2.98}}} & \multicolumn{1}{c} {\textbf{95.20{\scriptsize  $\pm$1.42}}} & {\textbf{98.95{\scriptsize  $\pm$0.36}}} \\ \bottomrule
	\end{tabular}
\end{table*}

\begin{figure} [!htb]
	\centering
	\includegraphics[scale=0.45]{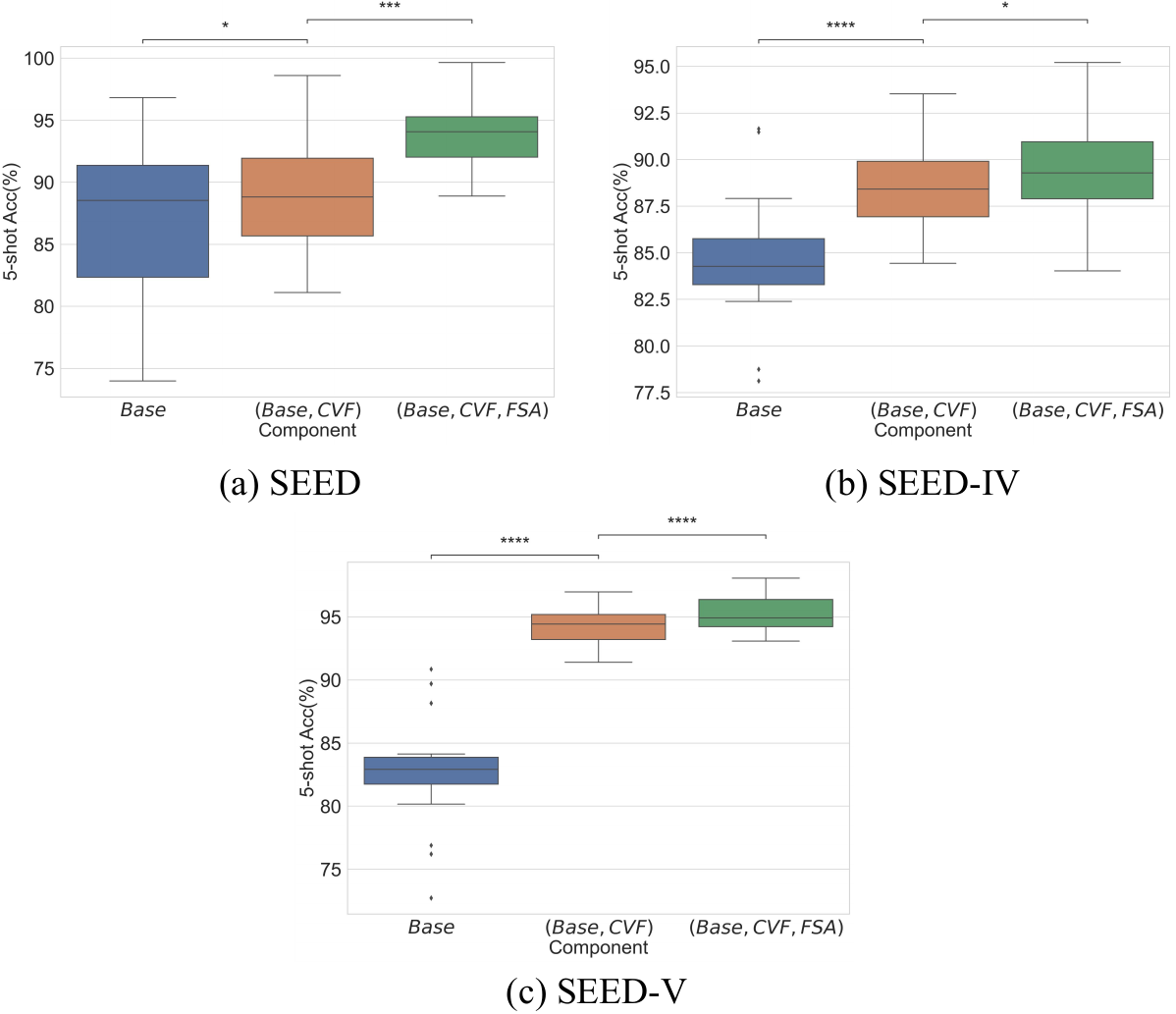}
	\caption{Statistical comparison of different components on the (a) SEED, (b) SEED-IV, and (c) SEED-V datasets. The asterisk ($\ast$) indicates $0.01\textless p\leq0.05$ while the four asterisks (****) denote $p\leq0.0001$, with a higher number of asterisks representing a more significant improvement.
	}
	\label{fig:t-test}
\end{figure}

The statistical analysis in Fig.~\ref{fig:t-test} employs pairwise t-tests to quantify module effectiveness under the 5-shot setting. Notably, the consistent presence of multiple asterisks across all datasets confirms that both CVF and FSA components contribute substantially to performance improvement, with p-values falling below $0.0001$ in most cases. This evidence highlights the effectiveness of the proposed module.

\subsection{Discussion on Impact of the CVF Module}
\label{sec:cvf}
We conduct additional experiments to rigorously validate the effectiveness of the CVF module. To isolate its effect, the FSA component is removed, and the CVF module is applied directly on the baseline. Building on this setting, we replaced the backbone encoders with several commonly used GCN-CNN combinations, including DGCNN~\cite{8320798}, Conv3, and Conv4~\cite{chen2019closerfewshot}. The observations in Table~\ref{table:cvf_db} confirm that CVF can be applied across diverse architectures while maintaining consistent performance gain on the GCN branch. At the same time, a performance drop rather than an additive improvement is observed in the DGCNN+Conv3 pairing in SEED-IV and SEED-V, which is likely related to the noticeable difference in learning stability between the two branches under the few-shot setting. Overall, these observations indicate that CVF is compatible with different backbone choices, while the extent of improvement depends on the balance and compatibility between the fused views.

\begin{table}[!htb]
	\centering
	\setlength{\tabcolsep}{2.5mm}
	\caption{Comparison of the CVF Module with Different Backbones on SEED, SEED-IV, and SEED-V}
	\begin{tabular}{lccc}
		\toprule
		& SEED & SEED-IV & SEED-V \\
		\midrule
		DGCN &{86.60{\scriptsize  $\pm$6.27}} &{84.65{\scriptsize  $\pm$3.75}} & {82.63{\scriptsize  $\pm$4.70}}\\
		Conv3 & {88.85{\scriptsize  $\pm$5.39}} &{80.74{\scriptsize  $\pm$6.32}} & {90.53{\scriptsize  $\pm$4.73}} \\
		CVF~(DGCN+Conv3) & \textbf{{93.70{\scriptsize  $\pm$3.85}}} &\textbf{{87.03{\scriptsize  $\pm$3.86}}} & \textbf{{91.08{\scriptsize  $\pm$3.38}}} \\
		\midrule
		DGCNN &{83.60{\scriptsize  $\pm$6.73}} &{62.75{\scriptsize  $\pm$6.96}} & {69.55{\scriptsize  $\pm$11.75}}\\
		Conv3 & {88.04{\scriptsize  $\pm$4.88}} &\textbf{{79.63{\scriptsize  $\pm$5.81}}} &\textbf{{90.41{\scriptsize  $\pm$4.64}}} \\
		CVF~(DGCNN+Conv3) &\textbf{{88.55{\scriptsize  $\pm$5.91}}} &{67.47{\scriptsize  $\pm$7.00}} & {87.63{\scriptsize  $\pm$6.14}} \\
		\midrule
		DGCNN &{83.02{\scriptsize  $\pm$7.15}} &{62.02{\scriptsize  $\pm$7.04}} & {69.16{\scriptsize  $\pm$11.50}}\\
		Conv4 & {88.50{\scriptsize  $\pm$6.36}} &{64.20{\scriptsize  $\pm$7.32}} & {82.15{\scriptsize  $\pm$7.99}} \\
		CVF~(DGCNN+Conv4) &\textbf{{89.20{\scriptsize  $\pm$4.90}}} &\textbf{{65.62{\scriptsize  $\pm$7.62}}} & \textbf{{82.87{\scriptsize  $\pm$8.58}}} \\
		\bottomrule
	\end{tabular}
	\label{table:cvf_db}
\end{table}

We further compare CVF with conventional attention mechanisms and two variants of the proposed method. CVF w/o MAML removes the meta-learning strategy, while $\mathrm{CVF}_{\mathrm{both}}$ applies residual refinement to both the graph and spatial views before concatenation. As shown in Table~\ref{table:cvf}, CVF consistently outperforms all comparison methods on the three datasets. The large performance gaps over self-attention and cross-attention show that directly applying conventional attention cannot fully capture the complementary information between the heterogeneous views. Furthermore, ablation studies reveal a non-negligible performance decline upon removing the meta-learning component implemented via MAML. This empirically validates its critical role in coordinating cross-view feature fusion. The lower performance of $\mathrm{CVF}_{\mathrm{both}}$ indicates that refining both views brings no additional benefit. This result is consistent with the motivation of CVF, where the spatial view is introduced to complement the graph representation and is therefore refined before fusion.

\begin{table}[!htb]
	\centering
	\setlength{\tabcolsep}{2.5mm}
	\caption{Comparison of the CVF Module with Different Fusion Strategies on SEED, SEED-IV, and SEED-V}
	\begin{tabular}{lccc}
		\toprule
		& SEED & SEED-IV & SEED-V \\
		\midrule
		Self Attention &{78.38{\scriptsize  $\pm$7.91}} &{70.59{\scriptsize  $\pm$7.04}} & {65.62{\scriptsize  $\pm$8.29}}\\
		Cross Attention & {71.91{\scriptsize  $\pm$9.83}} &{59.69{\scriptsize  $\pm$4.98}} & {73.17{\scriptsize  $\pm$9.70}} \\
		CVF w/o MAML & {93.28{\scriptsize  $\pm$4.10}} &{86.58{\scriptsize  $\pm$3.82}} & {90.52{\scriptsize  $\pm$3.42}} \\
		
		$\mathrm{CVF}_{\mathrm{both}}$ & {92.93{\scriptsize  $\pm$3.75}} & 		{86.15{\scriptsize  $\pm$3.40}} & {90.79{\scriptsize  $\pm$3.84}} \\
		\midrule
		CVF(ours) &\textbf{93.70{\scriptsize  $\pm$3.85}} & \textbf{87.03{\scriptsize  $\pm$3.86}} & \textbf{91.08{\scriptsize  $\pm$3.38}}\\
		\bottomrule
	\end{tabular}
	\label{table:cvf}
\end{table}

\begin{figure*} [t]
	\centering
	\includegraphics[scale=0.60]{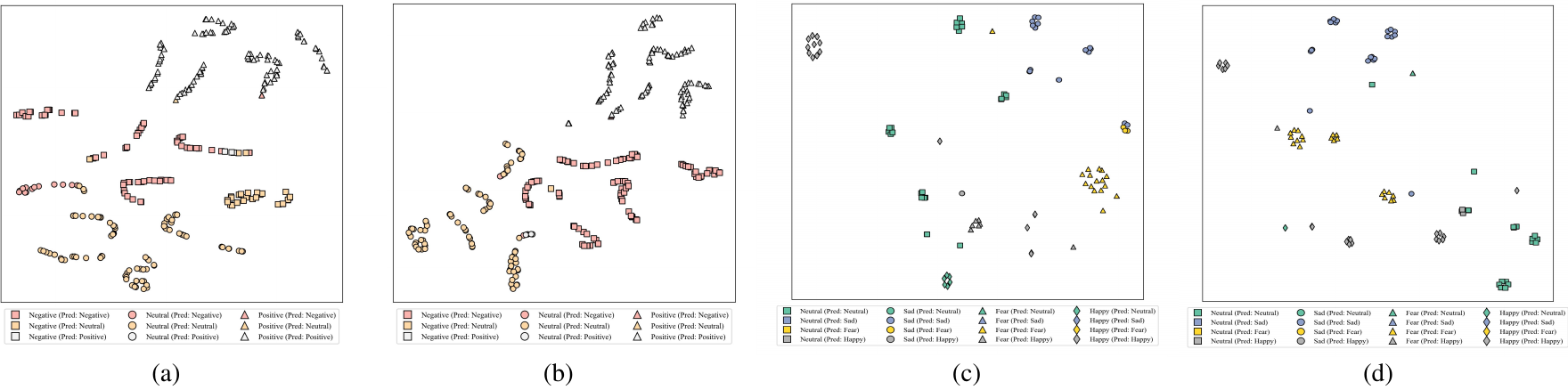}
	\caption{t-SNE visualization of the SEED and SEED-IV datasets with and without the CVF module under the 5-shot setting. (a) Baseline w/o CVF module on SEED, (b) Baseline w/ CVF module on SEED, (c) Baseline w/o CVF module on SEED-IV, and (d) Baseline w/ CVF module on SEED-IV. Markers represent the ground truth labels, while colors indicate the predicted labels.
	}
	\label{fig:tsne}
\end{figure*}

Fig.~\ref{fig:tsne} presents the t-SNE~\cite{van2008visualizing} visualization to illustrate the impact of the CVF module on feature representation. The model without the CVF module exhibits substantial inter-class overlap with ambiguous decision boundaries, indicating that the learned feature representations fail to fully distinguish between different emotions.  In contrast, integrating the CVF module results in more compact and well-separated clusters, suggesting that CVF enhances the model’s discriminative capability. This improvement is particularly evident in the SEED-IV dataset (Fig.~\ref{fig:tsne} (c) vs. Fig.~\ref{fig:tsne}(d)), where the cluster separation and alignment between ground truth markers and predicted colors become more pronounced. 

\subsection{Discussion on the Impact of the FSA Module}
\label{sec:fsa}
This subsection conducts comparative experiments and visual analysis to investigate the impact of the FSA module. In few-shot EEG emotion recognition, the support set can be viewed as an EEG prompt that guides the model toward subject-specific adaptation by conditioning the shared backbone on limited labeled examples. From this perspective, the FSA module plays a key role in enabling rapid and stable feature calibration for unseen subjects. To isolate its effect, we remove the CVF module and retain only the backbone encoder. Based on this simplified baseline, we replace the FSA module with several parameter-efficient methods~\cite{li2022cross,hu2022lora} commonly used in CV and NLP fields (e.g., Linear Probing, LoRA, and Adapter variants), ensuring the meta-learning strategy identical across all settings. As shown in Table~\ref{table:FSA}, our FSA module achieves the highest accuracy on SEED, SEED-IV, and SEED-V, demonstrating its superior ability to perform subject-specific calibration under few-shot conditions. The advantage of FSA mainly lies in its explicit adaptive bottleneck, which effectively captures subject-dependent feature shifts. In contrast, LoRA and other parameter-efficient methods only adjust low-rank or partial spaces of the backbone, limiting their capacity to model individualized variations in EEG patterns.

\begin{figure*} [t]
	\centering
	\includegraphics[scale=0.5]{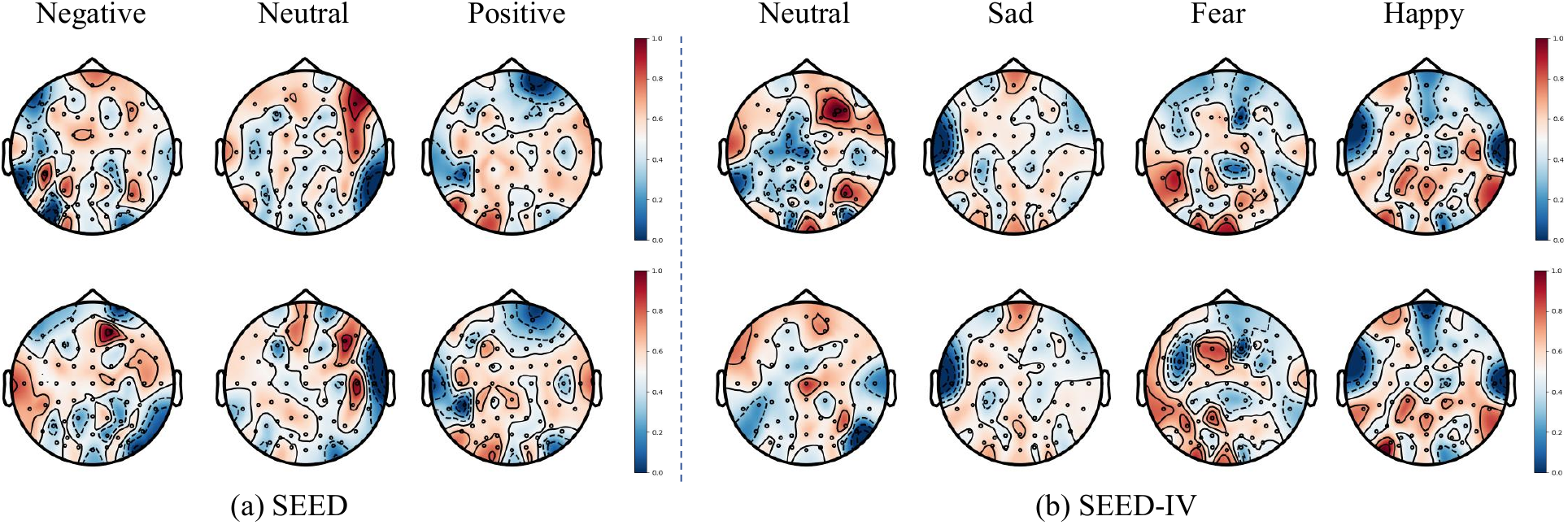}
	\caption{Brain topographic maps illustrating the effects of 5-shot adaptation on (a) the SEED and (b) the SEED-IV dataset. The first row depicts the state before adaptation, while the second row represents the state after adaptation.
	}
	\label{fig:brain_map}
\end{figure*}

Furthermore, this article visualizes the brain topographic maps before and after 5-shot adaptation using the FSA module, as illustrated in Fig.~\ref{fig:brain_map}. Specifically, this article employs the Shapley additive global importance~\cite{covert2020understanding} to quantify the importance of different channels under various emotional states. The fine-tuned model exhibits highly consistent and similar activated regions under the same emotions across different datasets. Meanwhile, for the negative emotions, the right hemisphere (particularly the right anterior/right prefrontal area) presents more prominent red or warm-colored regions. For neutral emotions, the overall brain region activation distribution appears relatively dispersed without conspicuous highly activated concentrated areas. In addition, positive emotions display stronger activation in the left hemisphere (especially in the left prefrontal lobe activity). These observations are broadly consistent with findings from neuroscience research~\cite{gold2015amygdala,kragel2016decoding,la2024effects}, thereby validating the effectiveness of the FSA module.

\begin{table}[!htb]
	\centering
	\setlength{\tabcolsep}{2.5mm}
	\caption{Comparison of the FSA Module with Different Parameter-efficient Methods on SEED, SEED-IV, and SEED-V}
	\begin{tabular}{lccc}
		\toprule
		& SEED & SEED-IV & SEED-V \\
		\midrule
		Linear Probing & {84.40{\scriptsize  $\pm$6.57}} &{84.24{\scriptsize  $\pm$4.03}} & {83.64{\scriptsize  $\pm$4.45}}\\
	    LoRA & 87.55{\scriptsize  $\pm$3.78} &83.14{\scriptsize  $\pm$3.57} & 79.29{\scriptsize  $\pm$5.62}\\
		Serial Adapter &{87.58{\scriptsize  $\pm$5.87}} &{85.11{\scriptsize  $\pm$4.13}} & {84.75{\scriptsize  $\pm$5.44}}\\
		Residual Adapter & {88.09{\scriptsize  $\pm$5.83}} &{86.96{\scriptsize  $\pm$3.46}} & {88.06{\scriptsize  $\pm$3.46}} \\
		\midrule
		FSA(ours) &\textbf{88.86{\scriptsize  $\pm$5.48}} & \textbf{88.75{\scriptsize  $\pm$2.74}} & \textbf{94.09{\scriptsize  $\pm$1.58}}\\
		\bottomrule
	\end{tabular}
	\label{table:FSA}
\end{table}

\subsection{Discussion on the Impact of Data Quantity and Diversity}
\label{sec:dqd}

\begin{figure}[!htb]
	\centering
	\begin{minipage}{0.49\linewidth}
		\centering
		\includegraphics[width=\linewidth]{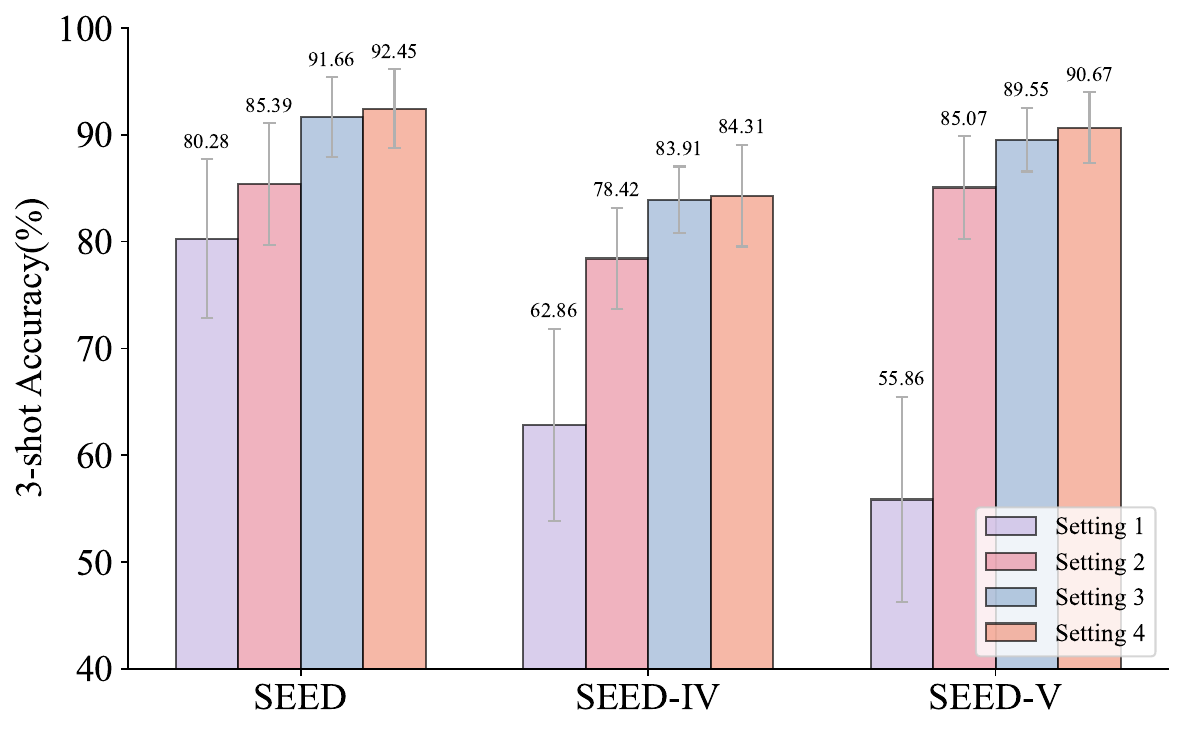}
	\end{minipage}
	\hfill
	\begin{minipage}{0.49\linewidth}
		\centering
		\includegraphics[width=\linewidth]{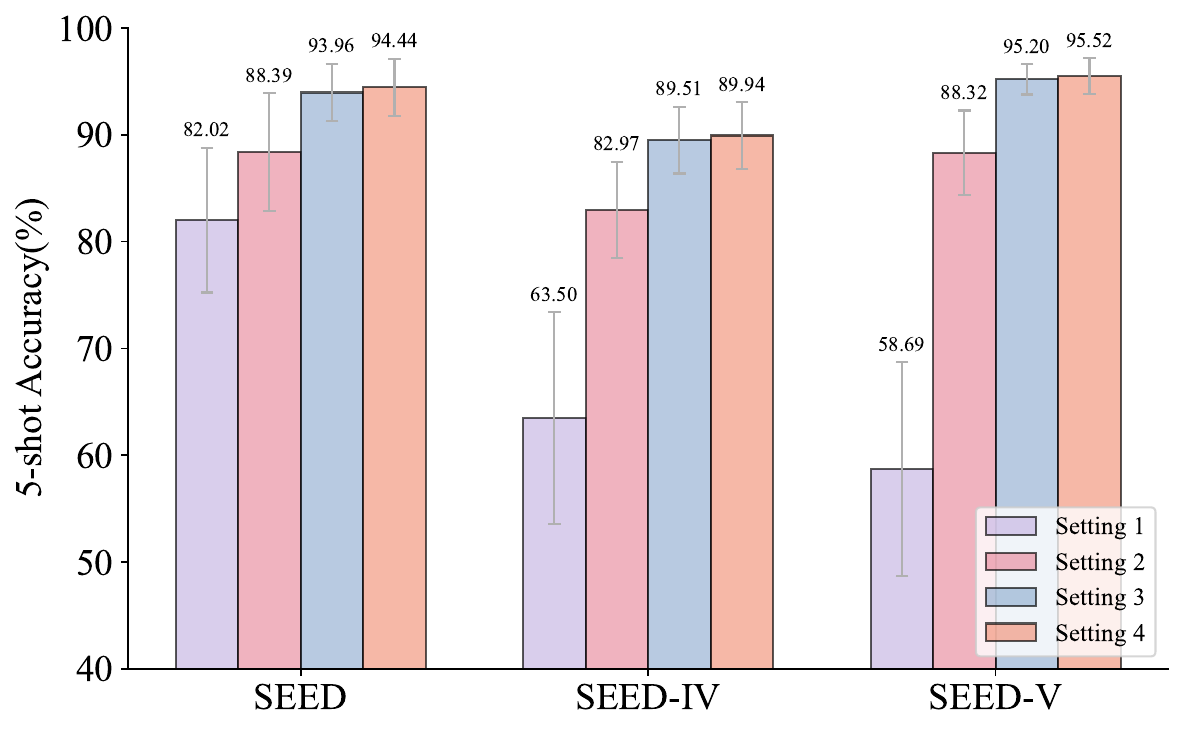}
		
	\end{minipage}
	\caption{3-shot and 5-shot results on SEED, SEED-IV, and SEED-V. Setting 1 draws adaptation samples from one trial per class and tests on the other trials. Setting 2 retains about half of the trials, Setting 3 uses the standard protocol, and Setting 4 uses all three sessions of the source subjects for training.}
	\label{fig:fs_ds}
\end{figure}

To examine the effects of data quantity and trial diversity, we evaluate FACE under four settings using both 3-shot and 5-shot setups. Setting 1 adopts a stricter trial-wise protocol, where $K$ adaptation samples are drawn from one trial per emotion, and testing is conducted on the other trials. Setting 2 reduces trial diversity by restricting both pre-training and adaptation to the first 2 of 5 trials per class in SEED, the first 3 of 6 in SEED-IV, and the first 2 of 3 in SEED-V. Setting 3 follows the standard cross-subject protocol, where one session of the target subject forms the target domain and the single-session data of the other subjects form the source domain. Setting 4 expands the source domain to all three sessions of the other subjects. Since Setting 1 uses a different and more challenging test set, it is reported as a conservative reference. As shown in Fig.~\ref{fig:fs_ds}, expanding the source data in Setting 4 provides performance gains over Setting 3, whereas reducing trial diversity in Setting 2 causes a larger decrease, showing the importance of exposure to different stimulus trials. Setting 1 yields lower accuracy because it introduces cross-trial shift in addition to subject shift. The decrease is more pronounced on SEED-IV and SEED-V, where finer-grained emotion classification makes the differences among classes more difficult to capture from a single adaptation trial. Moreover, increasing from 3-shot to 5-shot under Setting 1 provides limited gains because the additional samples come from the same stimulus and do not expand trial coverage.

\subsection{Discussion on the Impact of the Number of Shots}
\label{sec:num_shot}
\begin{figure} [!htb]
	\centering
	\includegraphics[scale=0.45]{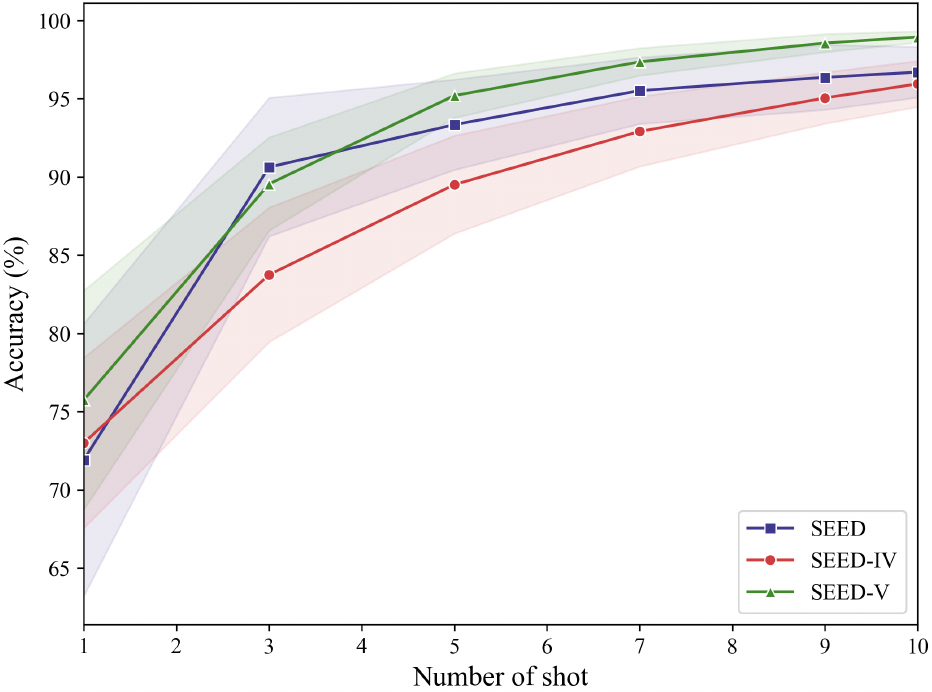}
	\caption{Accuracy Trends of SEED, SEED-IV and SEED-V across Different Shot Numbers
	}
	\label{fig:n_shot}
\end{figure}

Fig.~\ref{fig:n_shot} illustrates the accuracy trends of SEED, SEED-IV, and SEED-V under $\{1, 3, 5, 7, 9, 10\}$-shot settings. All three benchmarks demonstrate consistent improvements in classification accuracy as the number of shots increases. This upward trend suggests that more fine-tuning examples improve the models' ability to learn discriminative features. When the number of shots reaches 10, the accuracy across all three datasets reach accuracies ranging from 95.95\% to 98.95\%, indicating that providing 10 labeled samples for fine-tuning is sufficient to achieve promising performance. Moreover,  the shaded regions surrounding each performance curve quantify the standard deviation across multiple subjects. The SEED-V dataset achieves superior accuracy and exhibits relatively low variance, which may be attributed to its higher data quality. 

\subsection{Discussion on the Impact of Temporal Resolution}
\label{sec:seg_len}
The temporal resolution used during DE feature extraction plays a crucial role in determining both the stability and reliability of EEG representations. Notably, the SEED dataset provides DE features extracted with a 1-second window, whereas SEED-IV and SEED-V are officially released using 4-second segments. To examine the effect of temporal resolution, we additionally reprocessed SEED-IV and SEED-V using a unified 1-second window. The DE features were regenerated following the standard pipeline used in SEED, including artifact removal, down-sampling, feature extraction across five frequency bands, and LDS smoothing. We further evaluated our framework under this unified setting alongside recent representative methods~\cite{10509712,10160130}. As shown in Table~\ref{tab:main_results}, shorter 1-second segments result in a clear performance reduction across all approaches due to weaker temporal stability. Nevertheless, our framework remains consistently superior across datasets and exhibits a substantially smaller relative drop compared with other methods. These findings indicate that although the official 4-second windows of SEED-IV and SEED-V provide more stable DE features, the overall conclusions and comparative trends of this study remain robust under unified temporal segmentation.

\begin{table}[t]
	\centering
	\caption{Classification accuracy (\%) on SEED-IV and SEED-V datasets under 1s and 4s segment lengths.}
	\label{tab:main_results}
	\renewcommand{\arraystretch}{1.2}
	\setlength{\tabcolsep}{2.5pt}
	\begin{tabular}{lcccc}
		\toprule
		\textbf{Method} &
		\makecell{\textbf{SEED-IV}\\(1s)} &
		\makecell{\textbf{SEED-IV}\\(4s)} &
		\makecell{\textbf{SEED-V}\\(1s)} &
		\makecell{\textbf{SEED-V}\\(4s)} \\
		\midrule
		PR-PL~\cite{10160130}             & 81.32$\pm$8.53  & --            & --            & -- \\
		$\text{PR-PL}_{\text{avg}}^{*}$& 71.96$\pm$7.56  & 74.46$\pm$4.56 & 64.19$\pm$8.48  & 72.62$\pm$12.67 \\
		$\text{PR-PL}_{\text{max}}^{*}$ & 81.89$\pm$9.57  & 84.64$\pm$4.49 & 76.30$\pm$11.17 & 83.52$\pm$11.17 \\
		\midrule
		SSAA-MSDA~\cite{10509712}         & -- & 76.23$\pm$9.02 & -- & -- \\
		$\text{SSAA-MSDA}_{\text{avg}}^{*}$    & 67.91±6.47 & 73.65$\pm$6.73 & 53.69±6.09 &63.09$\pm$10.01 \\
		$\text{SSAA-MSDA}_{\text{max}}^{*}$         & 78.40±10.34 & 83.73$\pm$7.64 & 62.75±5.43 & 72.57$\pm$10.55 \\
		\midrule
		$\text{FACE}_{\text{avg}}(K=3)$      & 80.50$\pm$4.00 & 83.91$\pm$3.87 & 84.39$\pm$2.89 & 89.55$\pm$2.98 \\
		$\textbf{FACE}_{\text{avg}}(K=5)$ & \textbf{86.97$\pm$2.80} & \textbf{89.51$\pm$3.13} & \textbf{91.79$\pm$1.43} & \textbf{95.20$\pm$1.42} \\
		\bottomrule
		\multicolumn{5}{l}{\footnotesize \makecell[l]{1. $\ast$ denotes results reproduced from open-source implementation.
				\\2. avg: mean over sessions per subject, then average across subjects.
				\\ 3. max: select best session per subject, then average across subjects. }}
	\end{tabular}
\end{table}

\subsection{Parameter Sensitivity Analysis}
\label{sec:Para}
Table~\ref{table:para} presents the model performance with different numbers of attention heads ($L$) across various datasets and few-shot settings. As the number of attention heads varies from 1 to 5 across different shot conditions, the performance shows some fluctuations. However, these changes do not seem to follow a clear upward or downward trend directly related to the increase in the number of attention heads. Although there are differences between different settings, no significant pattern emerges indicating a strong impact on the number of attention heads. Consequently, these findings suggest that varying the number of attention heads does not produce significant performance differences across datasets and shot conditions.

\begin{table}[!htb]
	\centering
	\setlength{\tabcolsep}{2.5mm}
	\caption{Effect of the number of attention heads}
	\begin{tabular}{llccc}
		\toprule
		& &SEED & SEED-IV & SEED-V \\
		\midrule
		\multirow{3}{*}{$K=3$} &$L=1$ & {90.63{\scriptsize  $\pm$4.43}} &{83.75{\scriptsize  $\pm$4.30}} & \textbf{89.55{\scriptsize  $\pm$2.98}}\\
		&$L=2$ &\textbf{91.66{\scriptsize  $\pm$3.72}} &\textbf{83.91{\scriptsize  $\pm$3.87}} & {89.37{\scriptsize  $\pm$3.16}}\\
		&$L=5$ & {90.51{\scriptsize  $\pm$4.52}} &{83.89{\scriptsize  $\pm$4.26}} & {89.46{\scriptsize  $\pm$2.72}} \\
		\midrule
		\multirow{3}{*}{$K=5$} &$L=1$ & {93.34{\scriptsize  $\pm$2.89}} &\textbf{89.51{\scriptsize  $\pm$3.13}} & \textbf{95.20{\scriptsize  $\pm$1.42}}\\
		&$L=2$ &\textbf{93.96{\scriptsize  $\pm$2.70}} &{89.30{\scriptsize  $\pm$2.94}} & {95.09{\scriptsize  $\pm$1.39}}\\
		&$L=5$ & {93.50{\scriptsize  $\pm$3.17}} &{89.37{\scriptsize  $\pm$3.04}} & {95.16{\scriptsize  $\pm$1.26}} \\
		\midrule
		\multirow{3}{*}{$K=10$} &$L=1$ & {96.70{\scriptsize  $\pm$1.62}} &\textbf{95.95{\scriptsize  $\pm$1.47}} & {98.86{\scriptsize  $\pm$0.51}}\\
		&$L=2$ &\textbf{96.72{\scriptsize  $\pm$1.81}} &{95.87{\scriptsize  $\pm$1.25}} & {98.91{\scriptsize  $\pm$0.46}}\\
		&$L=5$ & {96.42{\scriptsize  $\pm$1.91}} &{95.88{\scriptsize  $\pm$1.39}} &\textbf{98.95{\scriptsize  $\pm$0.36}} \\
		\bottomrule
		\label{table:para}
	\end{tabular}
\end{table}

\section{Conclusion}
\label{conclusion}

In this article, we propose FACE, a cross-view fusion and few-shot adaptation framework for cross-subject EEG emotion recognition. Specifically, FACE first extracts global topological features through a dynamic graph convolutional branch and local spatial features through a convolutional branch. The two views are then mapped into a common feature space and fused to model their complementary relationships. Furthermore, a few-shot adapter is introduced to adjust the learned representation to an unseen subject using only a small number of labeled samples. We conduct experiments on three public EEG emotion datasets, namely SEED, SEED-IV, and SEED-V. The results show that FACE consistently outperforms existing methods under different few-shot settings. Despite the promising results, few-shot adaptation to unseen subjects remains challenging when labeled target samples lack sufficient diversity. Future work will investigate generative models for synthesizing diverse EEG samples tailored to unseen subjects and leverage multimodal information from diverse data sources~\cite{11457475,10210323} to improve robustness and generalization.

\bibliographystyle{IEEEtran}
\bibliography{ref}

\begin{thebibliography}{10}
\providecommand{\url}[1]{#1}
\csname url@samestyle\endcsname
\providecommand{\newblock}{\relax}
\providecommand{\bibinfo}[2]{#2}
\providecommand{\BIBentrySTDinterwordspacing}{\spaceskip=0pt\relax}
\providecommand{\BIBentryALTinterwordstretchfactor}{4}
\providecommand{\BIBentryALTinterwordspacing}{\spaceskip=\fontdimen2\font plus
\BIBentryALTinterwordstretchfactor\fontdimen3\font minus
  \fontdimen4\font\relax}
\providecommand{\BIBforeignlanguage}[2]{{%
\expandafter\ifx\csname l@#1\endcsname\relax
\typeout{** WARNING: IEEEtran.bst: No hyphenation pattern has been}%
\typeout{** loaded for the language `#1'. Using the pattern for}%
\typeout{** the default language instead.}%
\else
\language=\csname l@#1\endcsname
\fi
#2}}
\providecommand{\BIBdecl}{\relax}
\BIBdecl

\bibitem{lee2024encoding}
J.~P. Lee, H.~Jang, Y.~Jang, H.~Song, S.~Lee, P.~S. Lee, and J.~Kim, ``Encoding
  of multi-modal emotional information via personalized skin-integrated
  wireless facial interface,'' \emph{Nature Communications}, vol.~15, no.~1, p.
  530, 2024.

\bibitem{karl2024role}
V.~Karl, H.~Engen, D.~Beck, L.~B. Norbom, L.~Ferschmann, E.~R. Aksnes,
  R.~Kjelkenes, I.~Voldsbekk, O.~A. Andreassen, D.~Aln{\ae}s \emph{et~al.},
  ``The role of functional emotion circuits in distinct dimensions of
  psychopathology in youth,'' \emph{Translational Psychiatry}, vol.~14, no.~1,
  p. 317, 2024.

\bibitem{Li2022ACM}
X.~Li, Y.~Zhang, P.~Tiwari, D.~Song, B.~Hu, M.~Yang, Z.~Zhao, N.~Kumar, and
  P.~Marttinen, ``{EEG} based emotion recognition: A tutorial and review,''
  \emph{ACM Comput. Surv.}, vol.~55, no.~4, Nov. 2022.

\bibitem{10453943}
B.~Chen, C.~L.~P. Chen, and T.~Zhang, ``{GDDN}: Graph domain disentanglement
  network for generalizable {EEG} emotion recognition,'' \emph{IEEE
  Transactions on Affective Computing}, vol.~15, no.~3, pp. 1739--1753, 2024.

\bibitem{10609541}
M.~Wu, C.~L.~P. Chen, B.~Chen, and T.~Zhang, ``{Grop}: Graph orthogonal
  purification network for {EEG} emotion recognition,'' \emph{IEEE Transactions
  on Affective Computing}, pp. 1--14, 2024.

\bibitem{Xu_2025_ICCV}
Y.~Xu, B.~Liu, W.~Shao, Y.~Du, S.~He, and T.~Zhu, ``Cross-subject mind decoding
  from inaccurate representations,'' in \emph{Proceedings of the IEEE/CVF
  International Conference on Computer Vision (ICCV)}, October 2025, pp.
  15\,066--15\,075.

\bibitem{wilson2020survey}
G.~Wilson and D.~J. Cook, ``A survey of unsupervised deep domain adaptation,''
  \emph{ACM Transactions on Intelligent Systems and Technology (TIST)},
  vol.~11, no.~5, pp. 1--46, 2020.

\bibitem{9904850}
Z.~Li, E.~Zhu, M.~Jin, C.~Fan, H.~He, T.~Cai, and J.~Li, ``Dynamic domain
  adaptation for class-aware cross-subject and cross-session {EEG} emotion
  recognition,'' \emph{IEEE Journal of Biomedical and Health Informatics},
  vol.~26, no.~12, pp. 5964--5973, 2022.

\bibitem{10839595}
C.~Ahuja and D.~Sethia, ``{TRANSIT-EEG} — a framework for cross-subject
  classification with subject specific adaptation,'' \emph{IEEE Transactions on
  Cognitive and Developmental Systems}, pp. 1--16, 2025.

\bibitem{kouw2019review}
W.~M. Kouw and M.~Loog, ``A review of domain adaptation without target
  labels,'' \emph{IEEE transactions on pattern analysis and machine
  intelligence}, vol.~43, no.~3, pp. 766--785, 2019.

\bibitem{robey2021model}
A.~Robey, G.~J. Pappas, and H.~Hassani, ``Model-based domain generalization,''
  \emph{Advances in Neural Information Processing Systems}, vol.~34, pp.
  20\,210--20\,229, 2021.

\bibitem{khoee2024domain}
A.~G. Khoee, Y.~Yu, and R.~Feldt, ``Domain generalization through
  meta-learning: A survey,'' \emph{Artificial Intelligence Review}, vol.~57,
  no.~10, p. 285, 2024.

\bibitem{gharoun2024meta}
H.~Gharoun, F.~Momenifar, F.~Chen, and A.~H. Gandomi, ``Meta-learning
  approaches for few-shot learning: A survey of recent advances,'' \emph{ACM
  Computing Surveys}, vol.~56, no.~12, pp. 1--41, 2024.

\bibitem{10822213}
L.~Wang, J.~Zhu, L.~Du, B.~Jin, and X.~Wei, ``{FreAML}: A frequency-domain
  adaptive meta-learning framework for {EEG}-based emotion recognition,'' in
  \emph{2024 IEEE International Conference on Bioinformatics and Biomedicine
  (BIBM)}, 2024, pp. 1191--1198.

\bibitem{Chen_2025}
C.~Chen, H.~Fang, Y.~Yang, and Y.~Zhou, ``Model-agnostic meta-learning for
  {EEG}-based inter-subject emotion recognition,'' \emph{Journal of Neural
  Engineering}, vol.~22, no.~1, p. 016008, jan 2025.

\bibitem{ng2024subject}
H.~W. Ng and C.~Guan, ``Subject-independent meta-learning framework towards
  optimal training of {EEG}-based classifiers,'' \emph{Neural Networks}, vol.
  172, p. 106108, 2024.

\bibitem{10556575}
Z.~Wan, Q.~Yu, W.~Dai, S.~Li, and J.~Hong, ``Data generation for enhancing
  {EEG}-based emotion recognition: Extracting time-invariant and
  subject-invariant components with contrastive learning,'' \emph{IEEE
  Transactions on Consumer Electronics}, pp. 1--1, 2024.

\bibitem{9760385}
C.~Chen, Z.~Li, F.~Wan, L.~Xu, A.~Bezerianos, and H.~Wang, ``Fusing
  frequency-domain features and brain connectivity features for cross-subject
  emotion recognition,'' \emph{IEEE Transactions on Instrumentation and
  Measurement}, vol.~71, pp. 1--15, 2022.

\bibitem{10496191}
M.~Jin, C.~Du, H.~He, T.~Cai, and J.~Li, ``{PGCN}: Pyramidal graph
  convolutional network for {EEG} emotion recognition,'' \emph{IEEE
  Transactions on Multimedia}, vol.~26, pp. 9070--9082, 2024.

\bibitem{9765326}
Y.~Li, J.~Chen, F.~Li, B.~Fu, H.~Wu, Y.~Ji, Y.~Zhou, Y.~Niu, G.~Shi, and
  W.~Zheng, ``{GMSS}: Graph-based multi-task self-supervised learning for {EEG}
  emotion recognition,'' \emph{IEEE Transactions on Affective Computing},
  vol.~14, no.~3, pp. 2512--2525, 2023.

\bibitem{9751142}
Z.~Wang, Y.~Wang, J.~Zhang, C.~Hu, Z.~Yin, and Y.~Song, ``Spatial-temporal
  feature fusion neural network for {EEG}-based emotion recognition,''
  \emph{IEEE Transactions on Instrumentation and Measurement}, vol.~71, pp.
  1--12, 2022.

\bibitem{pessoa2017network}
L.~Pessoa, ``A network model of the emotional brain,'' \emph{Trends in
  cognitive sciences}, vol.~21, no.~5, pp. 357--371, 2017.

\bibitem{chen2018domain}
T.~Chen, B.~Becker, J.~Camilleri, L.~Wang, S.~Yu, S.~B. Eickhoff, and C.~Feng,
  ``A domain-general brain network underlying emotional and cognitive
  interference processing: evidence from coordinate-based and functional
  connectivity meta-analyses,'' \emph{Brain Structure and Function}, vol. 223,
  no.~8, pp. 3813--3840, 2018.

\bibitem{JAFARI2023107450}
M.~Jafari, A.~Shoeibi, M.~Khodatars, S.~Bagherzadeh, A.~Shalbaf, D.~L. García,
  J.~M. Gorriz, and U.~R. Acharya, ``Emotion recognition in {EEG} signals using
  deep learning methods: A review,'' \emph{Computers in Biology and Medicine},
  vol. 165, p. 107450, 2023.

\bibitem{houssein2022human}
E.~H. Houssein, A.~Hammad, and A.~A. Ali, ``Human emotion recognition from
  {EEG}-based brain--computer interface using machine learning: a comprehensive
  review,'' \emph{Neural Computing and Applications}, vol.~34, no.~15, pp.
  12\,527--12\,557, 2022.

\bibitem{song2020instance}
T.~Song, S.~Liu, W.~Zheng, Y.~Zong, and Z.~Cui, ``Instance-adaptive graph for
  {EEG} emotion recognition,'' in \emph{Proceedings of the AAAI Conference on
  Artificial Intelligence}, vol.~34, no.~03, 2020, pp. 2701--2708.

\bibitem{9857970}
M.~Sun, W.~Cui, S.~Yu, H.~Han, B.~Hu, and Y.~Li, ``A dual-branch dynamic graph
  convolution based adaptive transformer feature fusion network for {EEG}
  emotion recognition,'' \emph{IEEE Transactions on Affective Computing},
  vol.~13, no.~4, pp. 2218--2228, 2022.

\bibitem{10568943}
L.~Gong, W.~Chen, and D.~Zhang, ``An attention-based multi-domain bi-hemisphere
  discrepancy feature fusion model for {EEG} emotion recognition,'' \emph{IEEE
  Journal of Biomedical and Health Informatics}, vol.~28, no.~10, pp.
  5890--5903, 2024.

\bibitem{li2023novel}
J.~Li, F.~Wang, H.~Huang, F.~Qi, and J.~Pan, ``A novel semi-supervised meta
  learning method for subject-transfer brain--computer interface,''
  \emph{Neural Networks}, vol. 163, pp. 195--204, 2023.

\bibitem{10214058}
M.~Jiménez-Guarneros and G.~Fuentes-Pineda, ``Cross-subject {EEG}-based
  emotion recognition via semisupervised multisource joint distribution
  adaptation,'' \emph{IEEE Transactions on Instrumentation and Measurement},
  vol.~72, pp. 1--11, 2023.

\bibitem{sarafraz2024domain}
G.~Sarafraz, A.~Behnamnia, M.~Hosseinzadeh, A.~Balapour, A.~Meghrazi, and H.~R.
  Rabiee, ``Domain adaptation and generalization of functional medical data: A
  systematic survey of brain data,'' \emph{ACM Computing Surveys}, vol.~56,
  no.~10, pp. 1--39, 2024.

\bibitem{borgwardt2006integrating}
K.~M. Borgwardt, A.~Gretton, M.~J. Rasch, H.-P. Kriegel, B.~Sch{\"o}lkopf, and
  A.~J. Smola, ``Integrating structured biological data by kernel maximum mean
  discrepancy,'' \emph{Bioinformatics}, vol.~22, no.~14, pp. e49--e57, 2006.

\bibitem{chen2021ms}
H.~Chen, M.~Jin, Z.~Li, C.~Fan, J.~Li, and H.~He, ``{MS-MDA}: Multisource
  marginal distribution adaptation for cross-subject and cross-session {EEG}
  emotion recognition,'' \emph{Frontiers in Neuroscience}, vol.~15, p. 778488,
  2021.

\bibitem{10096469}
H.~Cai and J.~Pan, ``Two-phase prototypical contrastive domain generalization
  for cross-subject {EEG}-based emotion recognition,'' in \emph{ICASSP 2023 -
  2023 IEEE International Conference on Acoustics, Speech and Signal Processing
  (ICASSP)}, 2023, pp. 1--5.

\bibitem{10750375}
B.~Chen, C.~L.~P. Chen, and T.~Zhang, ``{Ugan}: Uncertainty-guided graph
  augmentation network for {EEG} emotion recognition,'' \emph{IEEE Transactions
  on Computational Social Systems}, pp. 1--13, 2024.

\bibitem{liu2024moge}
X.-H. Liu, W.-B. Jiang, W.-L. Zheng, and B.-L. Lu, ``{MoGE}: Mixture of graph
  experts for cross-subject emotion recognition via decomposing {EEG},'' in
  \emph{2024 IEEE International Conference on Bioinformatics and Biomedicine
  (BIBM)}.\hskip 1em plus 0.5em minus 0.4em\relax IEEE, 2024, pp. 3515--3520.

\bibitem{10856415}
B.~Liu, C.~Zheng, X.~Xu, C.~Xu, H.~Zhang, and S.~He, ``Rotation-adaptive point
  cloud domain generalization via intricate orientation learning,'' \emph{IEEE
  Transactions on Pattern Analysis and Machine Intelligence}, vol.~47, no.~5,
  pp. 4232--4239, 2025.

\bibitem{chen2021meta}
Y.~Chen, Z.~Liu, H.~Xu, T.~Darrell, and X.~Wang, ``Meta-baseline: Exploring
  simple meta-learning for few-shot learning,'' in \emph{Proceedings of the
  IEEE/CVF international conference on computer vision}, 2021, pp. 9062--9071.

\bibitem{wang2023improving}
L.~Wang, S.~Zhou, S.~Zhang, X.~Chu, H.~Chang, and W.~Zhu, ``Improving
  generalization of meta-learning with inverted regularization at
  inner-level,'' in \emph{Proceedings of the IEEE/CVF Conference on Computer
  Vision and Pattern Recognition}, 2023, pp. 7826--7835.

\bibitem{10445009}
H.~Liu, C.~L.~P. Chen, X.~Gong, and T.~Zhang, ``Robust saliency-aware
  distillation for few-shot fine-grained visual recognition,'' \emph{IEEE
  Transactions on Multimedia}, vol.~26, pp. 7529--7542, 2024.

\bibitem{snell2017prototypical}
J.~Snell, K.~Swersky, and R.~Zemel, ``Prototypical networks for few-shot
  learning,'' \emph{Advances in neural information processing systems},
  vol.~30, 2017.

\bibitem{ning2021cross}
R.~Ning, C.~P. Chen, and T.~Zhang, ``Cross-subject {EEG} emotion recognition
  using domain adaptive few-shot learning networks,'' in \emph{2021 IEEE
  international conference on bioinformatics and biomedicine (BIBM)}.\hskip 1em
  plus 0.5em minus 0.4em\relax IEEE, 2021, pp. 1468--1472.

\bibitem{9751421}
T.~Zhang, A.~E. Ali, A.~Hanjalic, and P.~Cesar, ``Few-shot learning for
  fine-grained emotion recognition using physiological signals,'' \emph{IEEE
  Transactions on Multimedia}, vol.~25, pp. 3773--3787, 2023.

\bibitem{10342627}
X.~Ning, J.~Wang, Y.~Lin, X.~Cai, H.~Chen, H.~Gou, X.~Li, and Z.~Jia,
  ``{MetaEmotionNet}: Spatial–spectral–temporal-based attention 3-d dense
  network with meta-learning for {EEG} emotion recognition,'' \emph{IEEE
  Transactions on Instrumentation and Measurement}, vol.~73, pp. 1--13, 2024.

\bibitem{finn2017model}
C.~Finn, P.~Abbeel, and S.~Levine, ``Model-agnostic meta-learning for fast
  adaptation of deep networks,'' in \emph{International conference on machine
  learning}.\hskip 1em plus 0.5em minus 0.4em\relax PMLR, 2017, pp. 1126--1135.

\bibitem{ye2024sg}
J.~Ye, Q.~Luo, J.~Yu, H.~Zhong, Z.~Zheng, C.~He, and W.~Li, ``{SG-BEV}:
  satellite-guided bev fusion for cross-view semantic segmentation,'' in
  \emph{Proceedings of the IEEE/CVF Conference on Computer Vision and Pattern
  Recognition}, 2024, pp. 27\,748--27\,757.

\bibitem{10384690}
S.~Chen, F.~Qin, X.~Ma, J.~Wei, Y.-T. Zhang, Y.~Zhang, and E.~Jovanov,
  ``Multi-view cross-fusion transformer based on kinetic features for
  non-invasive blood glucose measurement using ppg signal,'' \emph{IEEE Journal
  of Biomedical and Health Informatics}, vol.~28, no.~4, pp. 1982--1992, 2024.

\bibitem{li2022cross}
W.-H. Li, X.~Liu, and H.~Bilen, ``Cross-domain few-shot learning with
  task-specific adapters,'' in \emph{Proceedings of the IEEE/CVF conference on
  computer vision and pattern recognition}, 2022, pp. 7161--7170.

\bibitem{7104132}
W.-L. Zheng and B.-L. Lu, ``Investigating critical frequency bands and channels
  for {EEG}-based emotion recognition with deep neural networks,'' \emph{IEEE
  Transactions on Autonomous Mental Development}, vol.~7, no.~3, pp. 162--175,
  2015.

\bibitem{8283814}
W.-L. Zheng, W.~Liu, Y.~Lu, B.-L. Lu, and A.~Cichocki, ``{EmotionMeter}: A
  multimodal framework for recognizing human emotions,'' \emph{IEEE
  Transactions on Cybernetics}, vol.~49, no.~3, pp. 1110--1122, 2019.

\bibitem{liu2021comparing}
W.~Liu, J.-L. Qiu, W.-L. Zheng, and B.-L. Lu, ``Comparing recognition
  performance and robustness of multimodal deep learning models for multimodal
  emotion recognition,'' \emph{IEEE Transactions on Cognitive and Developmental
  Systems}, vol.~14, no.~2, pp. 715--729, 2022.

\bibitem{muller2019does}
R.~M{\"u}ller, S.~Kornblith, and G.~E. Hinton, ``When does label smoothing
  help?'' \emph{Advances in neural information processing systems}, vol.~32,
  2019.

\bibitem{10509712}
Y.~Yang, Z.~Wang, W.~Tao, X.~Liu, Z.~Jia, B.~Wang, and F.~Wan,
  ``Spectral-spatial attention alignment for multi-source domain adaptation in
  {EEG}-based emotion recognition,'' \emph{IEEE Transactions on Affective
  Computing}, vol.~15, no.~4, pp. 2012--2024, 2024.

\bibitem{10160130}
R.~Zhou, Z.~Zhang, H.~Fu, L.~Zhang, L.~Li, G.~Huang, F.~Li, X.~Yang, Y.~Dong,
  Y.-T. Zhang, and Z.~Liang, ``{PR-PL}: A novel prototypical representation
  based pairwise learning framework for emotion recognition using {EEG}
  signals,'' \emph{IEEE Transactions on Affective Computing}, vol.~15, no.~2,
  pp. 657--670, 2024.

\bibitem{li2023sparse}
\BIBentryALTinterwordspacing
B.~Li, Y.~Shen, J.~Yang, Y.~Wang, J.~Ren, T.~Che, J.~Zhang, and Z.~Liu,
  ``Sparse mixture-of-experts are domain generalizable learners,'' in \emph{The
  Eleventh International Conference on Learning Representations}, 2023.
  [Online]. Available: \url{https://openreview.net/forum?id=RecZ9nB9Q4}
\BIBentrySTDinterwordspacing

\bibitem{zhou2023progressive}
Y.~Zhou, F.~Li, Y.~Li, Y.~Ji, G.~Shi, W.~Zheng, L.~Zhang, Y.~Chen, and
  R.~Cheng, ``Progressive graph convolution network for eeg emotion
  recognition,'' \emph{Neurocomputing}, vol. 544, p. 126262, 2023.

\bibitem{9817639}
Y.~Peng, W.~Wang, W.~Kong, F.~Nie, B.-L. Lu, and A.~Cichocki, ``Joint feature
  adaptation and graph adaptive label propagation for cross-subject emotion
  recognition from {EEG} signals,'' \emph{IEEE Transactions on Affective
  Computing}, vol.~13, no.~4, pp. 1941--1958, 2022.

\bibitem{8320798}
T.~Song, W.~Zheng, P.~Song, and Z.~Cui, ``{EEG} emotion recognition using
  dynamical graph convolutional neural networks,'' \emph{IEEE Transactions on
  Affective Computing}, vol.~11, no.~3, pp. 532--541, 2020.

\bibitem{chen2019closerfewshot}
W.-Y. Chen, Y.-C. Liu, Z.~Kira, Y.-C. Wang, and J.-B. Huang, ``A closer look at
  few-shot classification,'' in \emph{International Conference on Learning
  Representations}, 2019.

\bibitem{van2008visualizing}
L.~Van~der Maaten and G.~Hinton, ``Visualizing data using t-sne.''
  \emph{Journal of machine learning research}, vol.~9, no.~11, 2008.

\bibitem{hu2022lora}
\BIBentryALTinterwordspacing
E.~J. Hu, yelong shen, P.~Wallis, Z.~Allen-Zhu, Y.~Li, S.~Wang, L.~Wang, and
  W.~Chen, ``Lo{RA}: Low-rank adaptation of large language models,'' in
  \emph{International Conference on Learning Representations}, 2022. [Online].
  Available: \url{https://openreview.net/forum?id=nZeVKeeFYf9}
\BIBentrySTDinterwordspacing

\bibitem{covert2020understanding}
I.~Covert, S.~M. Lundberg, and S.-I. Lee, ``Understanding global feature
  contributions with additive importance measures,'' \emph{Advances in Neural
  Information Processing Systems}, vol.~33, pp. 17\,212--17\,223, 2020.

\bibitem{gold2015amygdala}
A.~L. Gold, R.~A. Morey, and G.~McCarthy, ``Amygdala--prefrontal cortex
  functional connectivity during threat-induced anxiety and goal distraction,''
  \emph{Biological psychiatry}, vol.~77, no.~4, pp. 394--403, 2015.

\bibitem{kragel2016decoding}
P.~A. Kragel and K.~S. LaBar, ``Decoding the nature of emotion in the brain,''
  \emph{Trends in cognitive sciences}, vol.~20, no.~6, pp. 444--455, 2016.

\bibitem{la2024effects}
P.~La~Malva, A.~Di~Crosta, G.~Prete, I.~Ceccato, M.~Gatti, E.~D’Intino,
  L.~Tommasi, N.~Mammarella, R.~Palumbo, and A.~Di~Domenico, ``The effects of
  prefrontal tdcs and hf-trns on the processing of positive and negative
  emotions evoked by video clips in first-and third-person,'' \emph{Scientific
  Reports}, vol.~14, no.~1, p. 8064, 2024.

\bibitem{11457475}
H.~Liu, C.~L.~P. Chen, and T.~Zhang, ``Unified multimodal representation
  learning with prototype calibration for cross-subject few-shot emotion
  recognition,'' \emph{IEEE Transactions on Computational Social Systems},
  vol.~13, no.~3, pp. 2801--2817, 2026.

\bibitem{10210323}
X.~Gong, C.~L.~P. Chen, and T.~Zhang, ``Cross-cultural emotion recognition with
  {EEG} and eye movement signals based on multiple stacked broad learning
  system,'' \emph{IEEE Transactions on Computational Social Systems}, vol.~11,
  no.~2, pp. 2014--2025, 2024.

\end{thebibliography}
\begin{IEEEbiography}
	[{\includegraphics[width=1in,height=1.25in,clip,keepaspectratio]{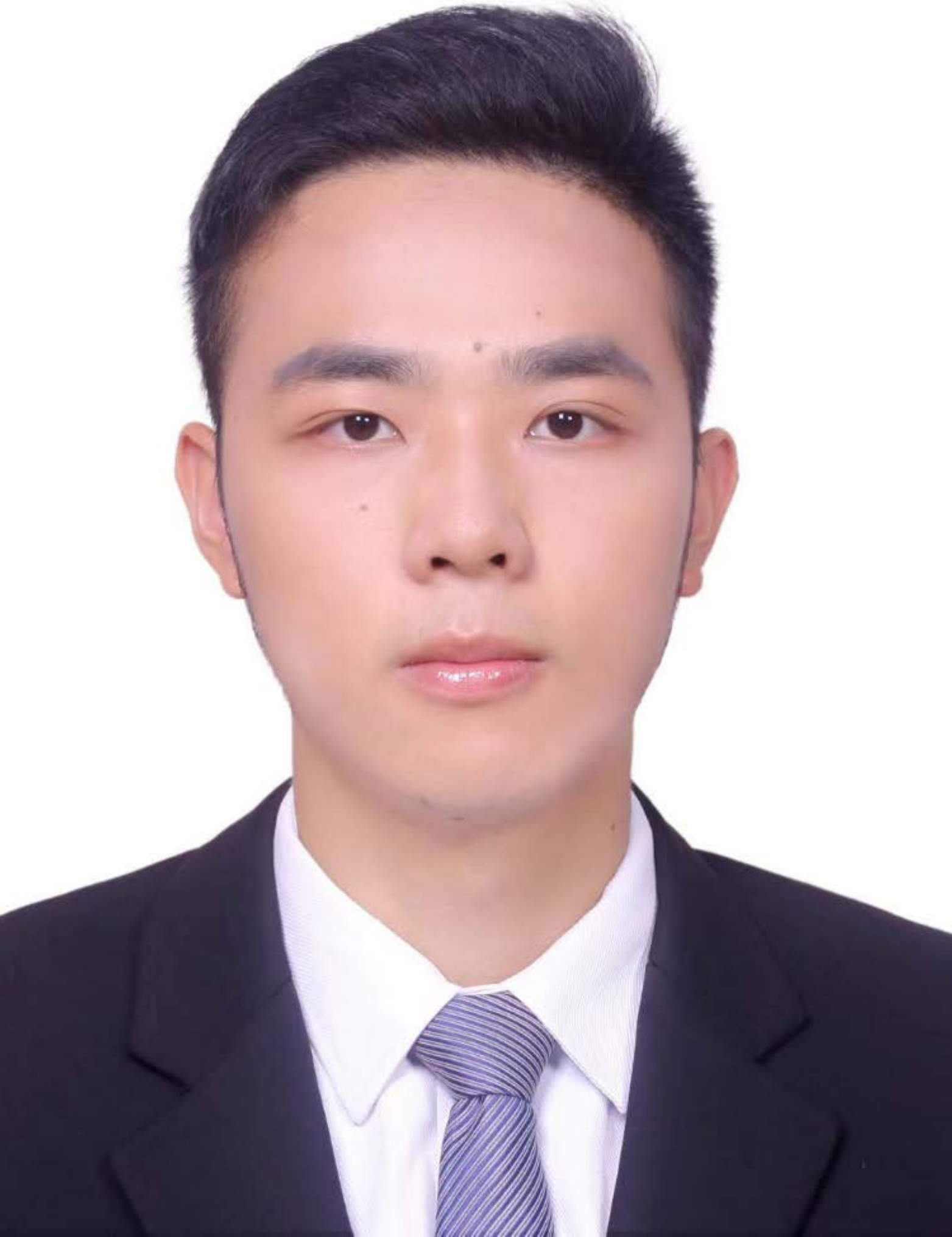}}] 
	{Haiqi Liu} received the B.S. and Ph.D. degrees in computer science and technology from South China University of Technology, Guangzhou, China, in 2021 and 2026, respectively. He is currently a postdoctoral fellow with the National University of Defense Technology, Changsha, China. His research interests include few-shot learning, image recognition, and affective computing.
\end{IEEEbiography}
\begin{IEEEbiography}
	[{\includegraphics[width=1in,height=1.25in,clip,keepaspectratio]{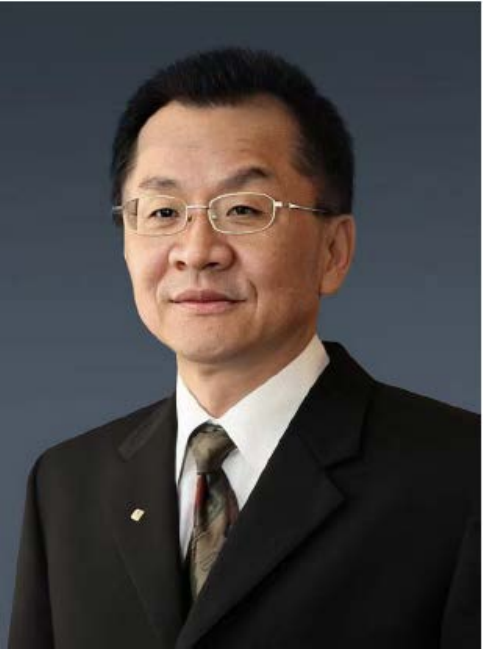}}] 
	{C. L. Philip Chen}(S’88–M’88–SM’94–F’07) received the M.S. degree from the University of Michigan at Ann Arbor, Ann Arbor, MI, USA, in 1985 and the Ph.D. degree from the Purdue University in 1988, all in electrical and computer science.
	
	He is the Chair Professor and Dean of the College of Computer Science and Engineering, South China University of Technology. He is the former Dean of the Faculty of Science and Technology. He is a Fellow of IEEE, AAAS, IAPR, CAA, and HKIE; a member of Academia Europaea (AE) and European Academy of Sciences and Arts (EASA). He received IEEE Norbert Wiener Award in 2018 for his contribution in systems and cybernetics, and machine learning. He is also a highly cited researcher by Clarivate Analytics in 2018, 2019, 2020, 2021 and 2022.
	
	He was the Editor-in-Chief of the IEEE Transactions on Cybernetics (2020-2021) after he completed his term as the Editor-in-Chief of the IEEE Transactions on Systems, Man, and Cybernetics: Systems (2014-2019), followed by serving as the IEEE Systems, Man, and Cybernetics Society President from 2012 to 2013. Currently, he serves as a deputy director of CAAI Transactions on AI, an Associate Editor of the IEEE Transactions on AI, IEEE Trans on SMC: Systems, and IEEE Transactions on Fuzzy Systems, an Associate Editor of China Sciences: Information Sciences. He received Macau FDCT Natural Science Award three times and a First-rank Guangdong Province Scientific and Technology Advancement Award in 2019. His current research interests include cybernetics, computational intelligence, and systems.
\end{IEEEbiography}

\begin{IEEEbiography}
	[{\includegraphics[width=1in,height=1.25in,clip,keepaspectratio]{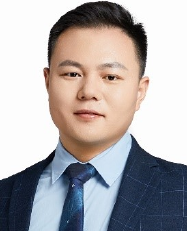}}] 
	{Tong Zhang} (M'16-SM’24) received the B.S. degree in software engineering from Sun Yat-sen University, at Guangzhou, China, in 2009, and the M.S. degree in applied mathematics from University of Macau, at Macau, China, in 2011, and the Ph.D. degree in software engineering from the University of Macau, at Macau, China in 2016. Dr. Zhang currently is a professor and Associate Dean of the School of Computer Science and Engineering, South China University of Technology, China. 
	
	His research interests include affective computing, evolutionary computation, neural network, and other machine learning techniques and their applications. Prof. Zhang is the Associate Editor of the IEEE Transactions on Affective Computing, IEEE Transactions on Computational Social Systems, and Journal of Intelligent Manufacturing. He has been working in publication matters for many IEEE conferences. 
\end{IEEEbiography}

\vfill
\end{document}